\documentclass[prb,two column,floats,showpacs,amsmath,amssymb]{revtex4}

\usepackage{dsfont}
\usepackage{graphicx}
\usepackage[caption=false]{subfig}
\usepackage[utf8x]{inputenc}
\usepackage[usenames,dvipsnames]{color}
\usepackage[shortlabels]{enumitem}
\usepackage{array}
\usepackage{hyperref}

\newcommand{\cL}{\mathcal{L}}
\newcommand{\cP}{\mathcal{P}}
\newcommand{\cQ}{\mathcal{Q}}

\newcommand{\oO}{\hat{O}}
\newcommand{\oH}{\hat{H}}
\newcommand{\oS}{\hat{S}}

\newcommand{\gp}{\hat{g}_{+}}
\newcommand{\gm}{\hat{g}_{-}}
\newcommand{\gpm}{\hat{g}_{\pm}}
\newcommand{\hp}{\hat{h}_{+}}
\newcommand{\hm}{\hat{h}_{-}}
\newcommand{\hpm}{\hat{h}_{\pm}}
\newcommand{\hmp}{\hat{h}_{\mp}}
\newcommand{\hz}{\hat{h}_{z}}

\newcommand{\lp}{\lambda_{+}}
\newcommand{\lm}{\lambda_{-}}
\newcommand{\lpm}{\lambda_{\pm}}
\newcommand{\lmp}{\lambda_{\mp}}
\newcommand{\lx}{\lambda_{x}}
\newcommand{\ly}{\lambda_{y}}
\newcommand{\lz}{\lambda_{z}}
\newcommand{\tlz}{\tilde{\lambda}_{z}}

\newcommand{\Gua}{\mathcal{G}_{\uparrow}}
\newcommand{\Gda}{\mathcal{G}_{\downarrow}}
\newcommand{\Gud}{\mathcal{G}_{\uparrow,\downarrow}}
\newcommand{\Fua}{\mathcal{F}_{\uparrow}}
\newcommand{\Fda}{\mathcal{F}_{\downarrow}}
\newcommand{\Fud}{\mathcal{F}_{\uparrow,\downarrow}}

\newcommand{\Gi}{\mathcal{G}_{i}}
\newcommand{\Fi}{\mathcal{F}_{i}}

\newcommand{\rs}{\hat{\rho}_{S}}
\newcommand{\ri}{\hat{\rho}_{I}}
\newcommand{\rt}{\hat{\rho}}

\newcommand{\rd}{R}

\newcommand{\ua}{\uparrow}
\newcommand{\da}{\downarrow}
\newcommand{\Tr}{\mbox{Tr}}

\newcommand{\cm}{c_{-}}
\newcommand{\cp}{c_{+}}
\newcommand{\cmp}{c_{\mp}}
\newcommand{\cpm}{c_{\pm}}

\newcommand{\on}{\hat{\omega}}

\newcommand{\ons}{\omega_{n}}

\newcommand{\Nti}{\tilde{N}}
\newcommand{\Dti}{\tilde{\Delta}}
\newcommand{\dti}{\tilde{\delta}}

\newcommand{\D}{\mathrm{d}}
\newcommand{\I}{\mathrm{i}}
\newcommand{\av}[1]{\big\langle #1 \big\rangle}
\newcommand{\bra}[1]{\left\langle #1 \right|}
\newcommand{\ket}[1]{\left| #1 \right\rangle}
\newcommand{\braket}[2]{\left\langle #1|#2 \right\rangle}

\newcommand{\ga}{$GaAs$}

\DeclareMathOperator*{\Res}{Res}
\DeclareMathOperator{\E}{e}

\newcommand{\Tab}[1]{Table \ref{#1}}

\begin{document}




\title{Spin decoherence in graphene quantum dots due to hyperfine interaction}

\author{Moritz Fuchs, Valentin Rychkov, and Bj\"orn Trauzettel}
\affiliation{Institute for Theoretical Physics and Astrophysics,
University of W\"urzburg, D-97074 W\"urzburg, Germany}

\date{\today}

\begin{abstract}
Carbon-based systems are prominent candidates for a solid-state spin-qubit due to weak spin-orbit and hyperfine interactions in combination with a low natural abundance of spin-carrying isotopes. We consider the effect of the hyperfine interaction on the coherence of an electron spin localized in a graphene quantum dot. It is known, that the hyperfine interaction in these systems is anisotropic promising interesting physics. We calculate the dynamics of an electron spin surrounded by a bath of nuclear spins in a non-Markovian approach using a generalized master equation. Considering a general form of the hyperfine interaction, we are able to extend the range of validity of our results to other systems beyond graphene. For large external magnetic fields, we find within Born approximation that the electron spin state is conserved up to small corrections, which oscillate with a frequency determined by the hyperfine interaction. The amplitude of these oscillations decays with a power law, where its initial value depends on the specific form of the anisotropy. Analyzing this in more detail, we identify two distinct classes of anisotropy, which can be both found in graphene depending on the orientation of the external magnetic field with respect to the carbon layer.
\end{abstract}


\maketitle

\section{Introduction}

Quantum dots (QDs) in solid state nanostructures have attracted a lot of interest in recent years, particularly, since localized spins are promising candidates\cite{Loss1998,Cerletti2005} for qubits in spin-based quantum computers. Most of these devices are realized in semi-conducting compounds of III-V materials implying an environment of many nuclear spins. In these structures, the hyperfine interaction (HI) between the nuclear spins and a central spin, for instance an electron or a hole spin, is a major source of both relaxation and decoherence of the central spin.\cite{Schliemann2003,Coish2009,Cywinski2011} The dynamics of these kind of spin systems has been extensively investigated both theoretically\cite{Schliemann2002,Khaetskii2002,Khaetskii2003,Coish2004,Klauser2006,Stepanenko2006,Giedke2006,Yao2006,Klauser2008,Coish2008,Fischer2008,Deng2008,Cywinski2009a,Testelin2009,Fischer2010,Coish2010,Barnes2011} and experimentally\cite{Hanson2007} covering the control of the nuclear environment\cite{Reilly2008,Xu2009,Latta2009,Vink2009,Gullans2010,Issler2010,Togan2011,Rudner2011,Carter2011} as well as the manipulation of the central spin.\cite{Greilich2009,Eble2009,Foletti2009,Ladd2010,Press2010,Bluhm2010,Bluhm2010a}
Since these investigations exhibit a decisive role of the HI, the use of materials with a low abundance of spin-carrying isotopes could offer a way to improve the properties of the qubits. 

Natural candidates for this are carbon and silicon based nanostructures, because the natural abundance of spin carrying $^{13}C$ and $^{29}Si$ is only about 1\% and 5\%, respectively. The experimental realization of carbon based QDs was achieved in recent years in graphene\cite{Stampfer2008,Ponomarenko2008,Schnez2009,Molitor2009,Guttinger2009,Liu2009,Molitor2010,Wang2010,Liu2010a,Guttinger2011,Fringes2011,Recher2010} as well as in carbon nanotubes\cite{Churchill2009,Chorley2011,Fang2011,Eichler2011}. In silicon, the qubits can be fabricated\cite{Morello2010,Morton2011} either using donor impurities or by confining a single electron via electro-statical gates. However, a controlled localization of the donor impurities is still a challenging task and electro-statically confined $Si$ QDs often involve nanostructures with other materials like $Ge$, which potentially introduce additional nuclear spins; see, for instance, Ref. \onlinecite{Morton2011} and references therein. 


Comparing the typical energy scale of the HI in different realizations of qubits reveals another advantage of carbon based systems. The hyperfine coupling constant\cite{Yazyev2008,Fischer2009a,Dora2010} $A_{^{13}C}$ is significantly smaller than in \ga{}, by about two orders of magnitude. Moreover, it is even less than the corresponding constant in $Si$ based systems, both donor impurities\cite{Ivey1975} and confined QDs,\cite{Assali2011} by approximately one order of magnitude. The same is true for \ga{} QDs using the spin of a heavy hole.\cite{Brunner2009,Testelin2009}

In this article, we will investigate the properties of an electron localized in a QD fabricated from a graphene sheet, where the HI is anisotropic due to the p-type nature of the electrons in the conduction band. This type of interaction has not been studied extensively so far, because the electronic properties of most qubits using the electron spin are primarily of the s-type and, hence, these systems are governed by an isotropic HI. Moreover, the few theoretical studies investigating anisotropic HI\cite{Khaetskii2002,Khaetskii2003} so far consider only a special case, where the longitudinal HI in the z direction deviates from the transverse interaction. Investigations of \ga{} qubits using the spin of a heavy-hole analyzed a totally anisotropic HI\cite{Fischer2008} or a HI, which is very weak in the transverse direction.\cite{Fischer2010} Finally, while most of the literature considers an isotropic HI in $Si$ systems, there are also studies\cite{Saikin2003,Witzel2007} taking the anisotropic HI into account. In contrast to these examples, we will analyze a system with a HI, which is in all spatial directions on the same order of magnitude. Experimentally, this physics is realized in a graphene QD, where the anisotropy can be varied by means of an external magnetic field.

This article is organized as follows: In Sec. \ref{sec:key_res}, the most important results are summarized. In Sec. \ref{sec:modham}, we investigate the physical properties of our system and its Hamiltonian in detail. This part is followed by a brief recapitulation of the Nakajima-Zwanzig master equation in Sec. \ref{sec:method}, which is used to investigate the dynamics of the electron spin in Sec. \ref{sec:results}. Finally, we conclude in section Sec. \ref{sec:conclusion}.

\section{Key results}
\label{sec:key_res}

\begin{figure}
\centering
\includegraphics[width=0.45\textwidth]{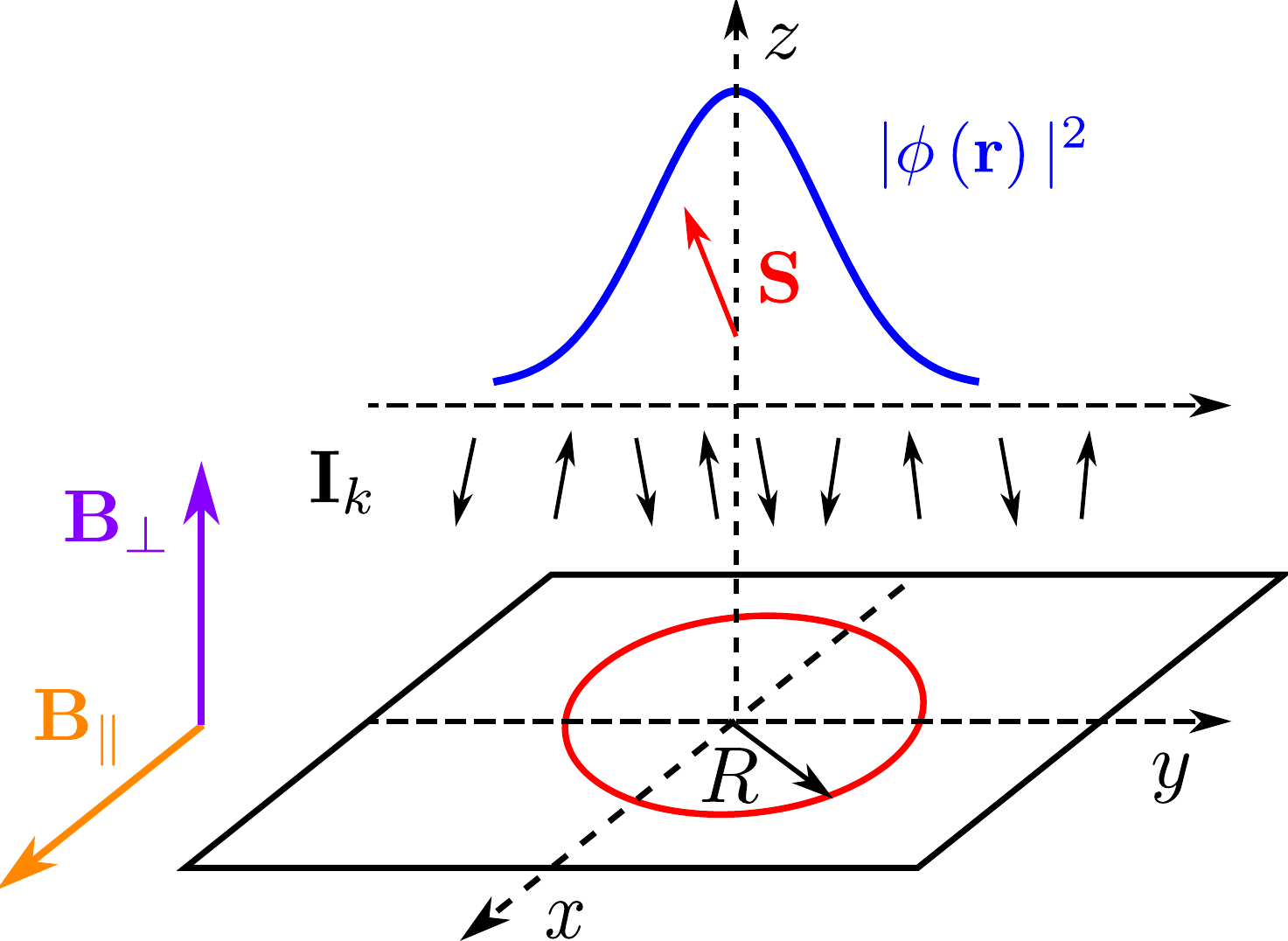}
\caption{(Color online) Schematic illustration of a two-dimensional quantum dot. An electron spin $\mathbf{S}$ is localized in a rotational symmetric QD, where it is in contact with bath of nuclear spins $\mathbf{I}_{k}$. Due to the confinement, the spatial distribution of the electron in the ground state of the QD may be described by a Gaussian envelope function $\phi(\mathbf{r})$ given in Eq. (\ref{eq:envfunc}), which in turn leads to a non-uniform HI between the electron spin and the nuclear spins. Finally, an external magnetic field is applied either perpendicular ($\mathbf{B}_{\perp}$) or parallel ($\mathbf{B}_{\parallel}$) with respect to the plane of the dot.}
\label{fig:schematic}
\end{figure}

We study the dynamics of an electron spin $\mathbf{S}$ in a graphene QD, where it is in contact with a bath of many nuclear spins $\mathbf{I}_{k}$  located at sites $\mathbf{r}_{k}$ via the HI as illustrated in Fig. \ref{fig:schematic}. Since there is no gap in the energy spectrum of graphene, the confinement of an electron is more challenging in contrast to electro-statically defined QDs in semi-conducting materials. One possibility to construct graphene nanostructures involves a chemical or mechanical treatment of graphene flakes, which has already been realized experimentally\cite{Stampfer2008,Ponomarenko2008,Schnez2009,Molitor2009,Guttinger2009,Liu2009,Molitor2010,Wang2010,Liu2010a,Guttinger2011,Fringes2011}. Furthermore, QDs can also be built by means of electrostatic potentials in presence of a finite energy gap, which can, for instance, be induced in single layer graphene by the substrate or in bilayer graphene by applying different potentials to the layers.\cite{Ohta2006,Castro2007,Zhang2009,Recher2009,Szafranek2010} 

We assume that the graphene flake is flat throughout the spatial extent of the QD, justifying our neglect of the influence of spin-orbit interaction on our problem. For simplicity, we consider further a rotational symmetric QD with the electron sitting in its ground state. Due to the confinement, the spatial probability distribution $|\phi(\mathbf{r})|^{2}$ of the electron is non-uniform and, thus, the HI becomes effectively site-dependent, since the major contribution to the coupling constant $A$ arises from on-site terms $A_{k}\propto A\cdot |\phi(r_{k})|^{2}$. For clarity, we omit constant factors and refer to Eq. (\ref{eq:coupconst}) for an exact definition of these couplings.

The total action of the nuclear spins can be interpreted as a nuclear magnetic field
\begin{equation}
\mathbf{h}=\sum_{k}A_{k}\mathbf{I}_{k}
\,,
\label{eq:nuc_mag}
\end{equation}
whose energy scale depends on the nuclear polarization $p$ and the hyperfine coupling constant: $\langle\hz\rangle\propto p\cdot A$. Additionally we allow for an external magnetic field $\mathbf{B}=(0,0,B_{z})$ giving rise to a Zeeman splitting $b_{S}=\hbar\gamma_{S}B_{z}$ of the electronic energy levels, where $\gamma_{S}\approx1.76\cdot10^{11}\,T^{-1}s^{-1}$ is the electron gyromagnetic ratio. This splitting is assumed to be much larger than the hyperfine energy: $b_{S}\gg A$. In principle the magnetic field can be arbitrary oriented with respect to the quantum dot plane. In this article, however, we restrict the orientation to be either perpendicular or parallel to this plane. Finally, we consider the system to be at zero temperature, which in particular implies that the thermal energy is much smaller than the spacing $\Delta E$ of the QD energy levels and the electron Zeeman energy: $k_{B}\,T\ll b_{S},\,\Delta E$.

This model is similar to previous studies of an electron spin confined to a \ga{}-QD,\cite{Coish2004} but there are, however, graphene-specific characteristics, which lead to new physics. Since the natural abundance $n_{I}$ of spin-carrying isotopes is small for carbon, only $N=n_{I}\cdot N_{tot}$ of all atoms $N_{tot}$ within the graphene dot carry spin, whereas all isotopes possess a spin in many semi-conducting materials like \ga{}, which is the most common material for building spin qubits. A comparison of the most important properties of graphene and \ga{} is given in \Tab{tab:13C/GaAs}. The HI coupling constant $A_{^{13}C}$ estimated\cite{Yazyev2008,Fischer2009a} for $n_{I}=1$, is about two orders of magnitude smaller than the constant $A_{GaAs}$ in \ga{} lowering the nuclear magnetic field by the same amount. For lower abundances the energy scale of this so-called Overhauser field is even further reduced by $\av{\hz}\propto A_{GaAs}\rightarrow n_{I}\cdot A_{^{13}C}$. In the following, we will explicitly distinguish between these two constants only if it is necessary to avoid confusion, while we use the general constant $A$ apart from that.  The comparably small hyperfine energy in graphene extends the typical time scale $\tau_{HI}\propto \hbar/A$ of this interaction significantly as compared to \ga{}. Moreover, the use of much weaker external magnetic fields $\mathbf{B}$ becomes possible in a graphene QD, which can be quantified analyzing the ratio of the HI coupling constant and the electron Zeeman energy:
\begin{equation}
\Delta
\sim
n_{I}A\,/\,b_{S}
\,,
\end{equation}
where we omitted factors of order $O(1)$ for brevity and clarity. The exact definition is postponed to Eq. (\ref{eq:Delta}) in Sec. IV, where this ratio will serve as a small parameter justifying our perturbative treatment of the HI. For a natural $^{13}C$-concentration of $n_{I}=0.01$, the Zeeman energy due to a comparably weak external magnetic field $|\mathbf{B}|=2.6\;mT$ still exceeds the hyperfine energy by more than two orders of magnitude: $\Delta\approx10^{-2}$.
\begin{center}
\begin{table}
\renewcommand{\arraystretch}{1.5}
\begin{ruledtabular}
\begin{tabular}{*4{>{$}c<{$}}}
{}					& \mathrm{units}	& \mathbf{GaAs}	& \mathbf{^{13}C}			\\
\hline
N_{tot}					& [1]			& 10^{6}		& 10^{5}			\\
n_{I}					& [1]			& 1			& 0 \sim 0.01 \sim 1 		\\
N=n_{I}\cdot N_{tot}			& [1] 			& 10^{6} 		& 10^{3} 			\\
A					& \left[\mu eV\right]	& 90			& 0.6				\\
B_{z}\gg |\lz n_{I}\,A/\hbar\gamma_{S}|	& [T]			& \gg 3.5  		& \gg n_{I}\cdot2.6\cdot 10^{-3}\\
\tau_{HI}\propto \hbar/A			& \left[\mu s\right]	& 1			& 100				\\
\\
\mathrm{type}\;\mathrm{of}\;\mathrm{HI}	& 			&(1,1,1)		& \perp:\; (-1/2,-1/2,1)	\\
(\lx,\ly,\lz)				& 			&			& \parallel:\; (1,-1/2,-1/2)	\\
\end{tabular}
\end{ruledtabular}
\caption{Comparison of the most important parameters of \ga{} and graphene. The total number of nuclei $N_{tot}$ is estimated for a QD of typical size $\rd=50\,nm$. While all nuclei in \ga{} carry spin, the abundance $n_{I}$ of $^{13}C$ can in principle be modified,  where the natural abundance is only $n_{I}=0.01$. The HI constant $A$ in \ga{} is about two orders of magnitude higher than in graphene (with $n_{I}=1$) demanding lower external magnetic fields $B_{z}$ and leading to a prolonged typical hyperfine time scale $\tau_{HI}$. The specific values of the anisotropy constants $\lambda_{i}$ in graphene depends on the orientation of the external magnetic field, while in \ga{} the HI is isotropic for electrons.}
\label{tab:13C/GaAs}
\end{table}
\end{center}

Besides this experimentally motivated advantages, the study of a graphene QD is of general interest, since it demands the investigation of an anisotropic HI, which arises from the p-type nature of the electrons:
\begin{equation}
H_{HI}
=
\lz {h}_{z} {S}_{z} + \lx {h}_{x}{S}_{x} + \ly {h}_{y}{S}_{y}
\label{eq:HHI}
\end{equation}
This anisotropy is expressed in terms of three coupling constants $\lx,\ly$, and $\lz$, where the isotropic limit is given by $\lx=\ly=\lz$ as it is, for instance, found for an electron in \ga{}. In graphene, this anisotropy is related to the geometry of the carbon plane and, hence, arises between the in-plane ($\bar{x},\bar{y}$) and out-of-plane ($\bar{z}$) components\cite{Fischer2009a}. Defining the quantization axis $z$ by the external magnetic field $\mathbf{B}$, the specific form of the anisotropy in graphene for different orientations is obtained by a geometric projection: $\lbrace \bar{x},\bar{y},\bar{z} \rbrace\rightarrow \lbrace x,y,z \rbrace$. This generally allows for a wide range of possible anisotropies. In this article, we focus on two special cases, where a magnetic field perpendicular to the graphene plane gives rise to 
$-2\lx=-2\ly=\lz=1$, while an in plane magnetic field generates another form of anisotropy: $\lx=-2\ly=-2\lz=1$.

Due to the large Zeeman splitting, real spin flips between the electron and an arbitrary nucleus are forbidden by energy conservation, since the resulting energy difference cannot be compensated by the hyperfine energy. Thus, only virtual spin flip-flops, where the electron spin is flipped back and forth, are allowed. 
If the characteristic time scale $\tau_{dd}$ for dipolar spin interactions within the nuclear bath is much longer than the typical time scale $\tau_{HI}$ of the HI, these virtual processes are the only source of change in the configuration of the nuclear spins. As a consequence of this, there is no randomizing effect within the nuclear bath itself and, hence, the system is in a non-Markovian regime rather than in a Markovian regime, where a stochastic nuclear magnetic field $\mathbf{h}$ would lead to a fast decay of the electron spin. However, even in a non-Markovian regime arbitrary states of the nuclear spin bath will in general lead to a fast decrease of the electron spin amplitude\cite{Schliemann2002,Schliemann2003}. Thus, the nuclear spins have to be prepared in a so-called narrowed state, which is an eigenstate of the nuclear magnetic field: $\hz\ket{n}=\av{\hz}_{n}\ket{n}$. Such a narrowing can in principle be obtained by a measurement of the nuclear magnetic field\cite{Klauser2006,Stepanenko2006,Giedke2006,Klauser2008} or by polarizing the bath to a high degree $p\sim1$, which is, however, hardly feasible in experiments. Hence, the experimentally most promising approaches are realized by pumping schemes\cite{Reilly2008,Xu2009,Latta2009,Vink2009,Greilich2009,Xu2009,Gullans2010,Issler2010,Togan2011,Rudner2011,Carter2011}, where the HI itself is employed to achieve such a state-narrowing.

Assuming a narrowed nuclear spin state, the non-Markovian dynamics of the electron spin can be analyzed for arbitrary anisotropy as defined in Eq. (\ref{eq:HHI}). The calculations are carried out by the use of the so-called Nakajima-Zwanzig equation\cite{Fick1990,Breuer2002,Coish2004,Barnes2011}, which provides an integro-differential equation for the electron subsystem only. This equation, however, can not be solved exactly, but demands for further simplifications, which are obtained by a second order expansion in the HI. This truncation of higher order terms, however, sets an upper bound to the time regime $t\ll\Delta^{-1}\tau_{HI}$, in which our predictions are valid. For a detailed description of this expansion, we refer to Section \ref{sec:method} and turn to the results of these calculations. 

The spin dynamics will be described by expectation values of the transverse and longitudinal components, $\av{S_{+}}(t)$ and $\av{S_{z}}(t)$, respectively, which allow the interpretation of our results in terms of Bloch-like equations of motion. 

The expectation value of the transverse electron spin component exhibits two oscillating contributions, which differ in both their amplitudes and frequencies:
\begin{equation}
\av{S_{+}}(t)=\sigma_{osc}(t)+\sigma^{pow}_{dec}(t)
\,.
\label{eq:S+_schematic_perp}
\end{equation}
The first term is given by
\begin{equation}
\sigma_{osc}(t)=  \frac{\av{S_{+}}_{0}}{1+(\lx^{2}+\ly^{2})/4\cdot\delta}\exp\Big[\I \frac{\ons}{\hbar} t\Big]
\,,
\label{eq:sig_osc}
\end{equation}
which describes a simple precession with a frequency determined by the Zeeman energy $\ons$ of the effective magnetic field. This field consists of the external magnetic field $B_{z}$ and the nuclear magnetic field $\hz$ pointing in the same direction. The second contribution oscillates with a much smaller frequency $\tau_{HI}^{-1}$, where the HI timescale $\tau_{HI}=2N\hbar/|\lz| n_{I} A$ is depending on both the HI energy scale $A$ and the absolute value of the longitudinal coupling constant $|\lz|$. Moreover, this second term is much smaller of   
order $\delta=\Delta^2/N\ll\Delta$ and exhibits a power-law decay of its amplitude. The reason for the smallness of this contribution is the large Zeeman-splitting, which makes the virtual flip-flop processes very unlikely due to the enormous energy cost of the electron spin flipping. For times $t\lesssim\tau_{HI}$ only a numerical solution for $\sigma^{pow}_{dec}(t)$ was within reach, whereas for times $\tau_{HI}\lesssim t \ll \Delta^{-1} \tau_{HI}$ this function is asymptotically described by:
\begin{equation}
\sigma^{pow}_{dec}(t)=\delta\frac{\tau_{HI}}{t}\left(U\sin(\frac{t}{\tau_{HI}}) - \I V\cos(\frac{t}{\tau_{HI}})\right)
\,.
\label{eq:sig^pow_dec}
\end{equation}
The constants $U$ and $V$ are determined by the initial value $\av{S_{+}}_{0}=\av{S_{x}}_{0}+\I\av{S_{y}}_{0}$, the polarization $p$ of the nuclear bath and the couplings in the transverse plane, $\lx$ and $\ly$, respectively:
\begin{equation}
U
=
\ly^{2}\av{S_{x}}_{0} +\I \lx^{2}\av{S_{y}}_{0}
\label{eq:U}
\end{equation}
and
\begin{equation}
V
=
p\cdot\Big(\lx^{2}\av{S_{x}}_{0} +\I \ly^{2}\av{S_{y}}_{0}\Big)
\,.
\label{eq:V}
\end{equation}
These results are true for $\lz>0$, from which one obtains the time evolution of the electron spin for $\lz<0$ by applying the following relation: $(-|\lz|,\omega)\leftrightarrow(|\lz|,-\omega)$. Thus a negative coupling in longitudinal direction can be treated as an inversion of the effective nuclear magnetic field.
From Eqs. (\ref{eq:U}) and (\ref{eq:V}), we see, that there are two distinct classes of anisotropy, which differ in the amplitude of the decaying contribution $\sigma^{pow}_{dec}(t)$. For the first class, defined by $\lx=\ly$, the constants $U$ and $V$ are both proportional to the initial value of the transverse spin $\av{S_{+}}_{0}=\av{S_{x}}_{0}+\I\av{S_{y}}_{0}$. The reason for this is the preserved rotational symmetry in the transverse plane, which is for instance realized in a \ga{} QD with $\lx=\ly=\lz=1$ and in a graphene QD subjected to a perpendicular magnetic field with $-2\lx=-2\ly=\lz=1$. Due to this symmetry, there is no precisely defined direction within this plane distinguishing between $\av{S_{x}}_{0}$ and $\av{S_{y}}_{0}$. As a consequence of this, only their superposition $\av{S_{+}}_{0}=\av{S_{x}}_{0}+\I\av{S_{y}}_{0}$ matters for the initial state preparation.

This is, however, not the case for the second class of anisotropy defined by $\lx\neq\ly$, which is for example found in a graphene QD with in-plane magnetic field, where the couplings are given by $\lx=-2\ly=-2\lz=1$. Now, the constants $U$ and $V$ describe a weighted mixture of $\av{S_{x}}_{0}$ and $\av{S_{y}}_{0}$, which is caused by the broken rotational symmetry. This fact, in turn, allows to precisely define the $x$- and $y$-direction, with respect to which the transverse electron spin component can be prepared. A more detailed discussion of this is given at the end of section V.B.

In contrast to this, the longitudinal spin component does not show qualitative differences between the two classes. Its time evolution consists of two parts:
\begin{equation}
\av{S_{z}}(t)=\bar{\av{S_{z}}}+\sigma^{pow}_{dec}(t)\cdot\exp\Big[-\I \frac{\ons}{\hbar} t\Big]
\,.
\label{eq:Sz_schematic}
\end{equation}
This expectation value is dominated by a constant contribution $\bar{\av{S_{z}}}$, which is, up to small correction of order $O(\delta)$, given by the initial value $\bar{\av{S_{z}}}\sim\av{S_{z}}_{0}$. The second term behaves similar to the decaying part of the transverse spin component $\sigma^{pow}_{dec}(t)$ except for an additional overlying oscillation with frequency $\ons$. More importantly the constants
\begin{equation}
U=\Big[\av{S_{z}}_{0} - p\frac{\lx\ly}{\lx^2+\ly^2}\Big] \cdot \frac{\lx^2+\ly^2}{2}
\end{equation}
and
\begin{equation}
V=\Big[p\cdot\av{S_{z}}_{0} + \frac{\lx\ly}{\lx^2+\ly^2} \Big] \cdot \frac{\lx^2+\ly^2}{2}
\end{equation}
show only a quantitative dependence on the anisotropy, while they are qualitatively unchanged for arbitrary anisotropy. Thus, the longitudinal spin component shows the same qualitative behavior for any reasonable values of $\lx$ and $\ly$. 

Finally, we focus on the implications of our general results on the two graphene models, where the external magnetic field is either perpendicular ($\perp$) or parallel ($\parallel$) with respect to the carbon layer. While the former case shows the same qualitative behavior as an isotropic system, one faces a mixing of the initial transverse amplitudes due to the broken rotational symmetry for the later case. Moreover, as we have shown above, the time regime for which we find only a small partial decay is determined by the time scale of the HI. To be more precise, most of the electron spin is preserved for times $t\ll\Delta^{-1}\tau_{HI}$. Since this timescale depends on the absolute value of the coupling constant $|\lz|$ in the longitudinal direction, one observes a different behavior for perpendicular and parallel magnetic field, which is quantitatively described by the following ratio:
\begin{equation}
\frac{\tau_{HI}^{\parallel}}{\tau_{HI}^{\perp}}=\frac{|\lz^{\perp}|}{|\lz^{\parallel}|}=2
\,.
\end{equation}
Thus, the time regime, on which only a small partial power law decay of the spin amplitude and, vice versa, a mostly preserved electron spin is to be expected, is prolonged by a factor of $2$ for a parallel magnetic compared to a perpendicular oriented field. Therefore, we conclude, that a parallel orientation is preferable, although one has to prepare the transverse electron spin more carefully at the beginning of an experiment. 

\section{Model and Hamiltonian}
\label{sec:modham}

\subsection{Physical properties of the model}
\label{subsec:model}

In this section, we discuss some of the physical properties of the model introduced at the beginning of Sec. \ref{sec:key_res} in more detail. For brevity and clarity, we measure all spins in units of $\hbar$ and adjust the coupling constants to give the right dimensions. 

The quantum dot hosting the electron is defined by electro-statical gates on top of a graphene sheet, where we assume a rotational symmetric QD for simplicity. Due to the confinement, the QD has a discrete spectrum of bound states, with an energy splitting between different states\cite{Silvestrov2007,DeMartino2007,Trauzettel2007,Matulis2008,Recher2009,Titov2010}. If the temperature is small compared to the level-spacing $\Delta E$ of these bound state energies, the electron will occupy the ground state, which we describe by an envelope-function of the form:
\begin{equation}
\phi\left(\textbf{r}\right) 
=
\frac{1}{\sqrt{\pi}\rd}\exp\left[-\frac{1}{2}\left( \frac{r}{\rd}\right)^{2}\right]
\text{,}
\label{eq:envfunc}
\end{equation}
where $r=\left|\textbf{r}\right|$ is the absolute value of the electron position. This wave-function is a Gaussian with a Bohr radius $\rd$, which we will regard as the radius of our QD. Note that the envelope function is not the exact electron wave-function, but should give a good approximation to the precise solution. This can be seen, for instance, in graphene QDs based on semi-conducting armchair nano-ribbons\cite{Trauzettel2007}. The most important aspects, which are captured by this specific choice are the absence of nodes in the ground-state, a peak of the wave-function in the center as well as a strong decay inside the barriers. A QD with radius $\rd$ includes
\begin{equation}
N_{tot}
=
\pi \rd^{2}/V_{0}
\label{eq:ntot}
\end{equation}
nuclei in total, where $V_{0}=\sqrt{3}\,a^2/4$ is the (two-dimensional) unit cell volume containing one nucleus. For a typical graphene QD of size $R=50\,nm$ and the graphene lattice constant given by $a=2.46\,\mathring{A}$, there are $N_{tot}\sim10^{5}$ nuclei within the dot.

Naturally only a fraction $n_{I}\approx0.01$ of these nuclei is of the spin-carrying species $^{13}C$, while the remainder consists of spinless $^{12}C$. However, it should in principle be possible to modify this natural abundance by either isotopic purification or $^{13}C$ enrichment, as it was already done for other carbon systems\cite{Banholzer1992,Simon2005,Churchill2009}. This leads to a fraction $n_{I}=N/N_{tot}$ of $^{13}C$-atoms carrying spin $I=1/2$, where $0<n_{I}<1$. Especially the lower bound, where the number of nuclei becomes too small to treat them as a bath, is an interesting limit, which is, however, beyond the scope of this article. Besides the $^{13}C$ spins, we neglect all other possible sources of nuclear spins stemming from the gate- and substrate-materials, which is justified by the strong confinement of the electrons into the graphene plane. Altogether, we consider $N=n_{I}N_{tot}$ nuclear spins $I=1/2$ within the QD radius $\rd$ forming a spin-bath with an arbitrary polarization $p$, where $p=0$ for an unpolarized bath and $p=1$ for full polarization.

We first turn to the interaction between a single nuclear spin $\textbf{I}_{k}=\left(I_{k,x},I_{k,y},I_{k,z}\right)$, located at an arbitrary position $\textbf{r}_{k}$ within the QD, and the electron spin $\textbf{S}$ at position $\textbf{r}$. In general, the HI consists of three major contributions\cite{Stoneha1985,Abragam1989}: the Fermi contact interaction, the anisotropic HI, and the coupling of the electron orbital angular momentum to the nuclear spins. For a carbon nanotube one would have to take into account all three terms due to the curvature dependent mixture of $s$- and $p$-type wave-functions, while for a perfectly flat sheet of graphene only the anisotropic HI is relevant\cite{Fischer2009a}, where the anisotropy can be formulated in terms of coupling constants $\lambda_{i}$ and the strength of the HI is given by another constant $A$. This anisotropy arises from the p-type nature of the electron wave function and is, thus, related to the geometry of the graphene plane, where the in-plane and out of plane components deviate. Thus, the definition of the quantization axis via an external magnetic field allows us to create different forms of anisotropy as stated in Sec. \ref{sec:key_res}.

Furthermore, the HI also depends on the position of the nucleus due to the non-uniform spatial distribution of the electron within the QD, since the HI is dominated by on-site contributions, where the electron and the nucleus are located at the same site. Thus, the interaction $A_{k}$ at site $k$ is determined by the hyperfine coupling constant $A$ and the electron envelope function at $\mathbf{r}_{k}$:
\begin{equation}
A_{k} = A V_{0} \left|\phi\left(\textbf{r}_{k}\right)\right|^{2}
\,,
\label{eq:coupconst}
\end{equation}
where $V_{0}$ is the volume of a primitive unit cell. For a fully $^{13}C$ enriched system $n_{I}=1$ the hyperfine constant is \cite{Fischer2009a} $A=0.6\,\mu eV$ implying that a single scattering process is on the order of $A/N_{tot}=n_{I}\,A/N\,$.\cite{foot1} If we relate the radial position $r_{k}<\rd$ of an arbitrary nuclear spin to the number of nuclear spins $k$ within this radius $r_{k}$ by 
\begin{equation}
\left(\frac{r_{k}}{\rd}\right)^2=\frac{k}{N}
\text{.}
\label{eq:rk}
\end{equation}
and use Eq. (\ref{eq:envfunc}), we can rewrite the expression of the HI:
\begin{equation}
A_{k}
=
\frac{n_{I} A}{N}\exp\left[-\frac{k}{N}\right]
\label{eq:coupconst2}
\end{equation}
for a system of $N=n_{I}N_{tot}$ nuclear spins.

One important degree of freedom of an electron in graphene has so far not been taken into account. The model used here, does not include the valley degree of freedom. The valleys correspond to the two inequivalent Dirac cones of the hexagonal Brillouin zone and must be considered in scattering processes involving a short-range potential. As a consequence of this, one has to check, if the $r^{-3}$ dependency of the anisotropic HI is sufficient to generate valley scattering. Therefore, we additionally investigate the momentum dependence of the coupling constants $A=A\left(\textbf{Q}\right)$, where $\textbf{Q}$ is the difference between incoming and outgoing momentum. Assuming, that the most important contributions arise from on-site terms \cite{Fischer2009a}, we estimate, that the valley mixing $\textbf{Q}=\Delta\textbf{K}=\textbf{K}-\textbf{K}'$ and valley conserving $\textbf{Q}=0$ processes are of comparable strength.

This result reflects the short-range nature of the anisotropic HI leading to a broad distribution in momentum space. Hence, the valley scattering is relevant, unless it is forbidden due to energy conservation, which demands that the HI compensates the energy splitting of the two valleys\cite{Palyi2009}. From Eq. (\ref{eq:coupconst2}), we know, that the energy scale of scattering processes between the electron and a single nuclear spin is on average of order $\epsilon_{HI}\propto n_{I} A/N\approx10^{-12}\,eV$. In typical graphene QDs, we expect a non-zero valley splitting $\Delta_{\textbf{K,\textbf{K}'}}$ because of several mechanisms. First, every experimental setup will have deviations from an ``ideal`` setup, e.g. roughness of the boundaries or adatoms, which couple the valleys. But even in an ideal experiment, one can expect a valley-splitting due to the presence of an external magnetic-field in combination with a finite mass term induced by the substrate\cite{Recher2009}. Furthermore, if we assume a QD made of a semi-conducting graphene ribbon with armchair boundaries, the valley degeneracy is also lifted by $\frac{\hbar\,\pi v_{F}}{3 w}\sim10\,meV$, where $w=50\,nm$ is a typical width of a ribbon and $v_{F}=10^{6}\,m/s$ is the Fermi-velocity in graphene.

Finally, let us briefly comment on the dynamics within the nuclear spin bath. This internal dynamics would be induced by nuclear dipolar interaction, which is, however, suppressed in QDs due to the ''frozen core'' effect\cite{Ramanathan2008,Coish2010} and, hence, negligible. As a consequence of this, we assume that the dynamics of the bath can be neglected on time scales $t\sim\tau_{HI}$ relevant for the electron spin evolution.

\subsection{Hamiltonian}
\label{subsec:hamiltonian}

We start this paragraph by formalizing the most important facts of the two previous sections, where we have discussed the physical properties of our model in detail. Note that we replace classical variables $\mathbf{S},\mathbf{I},\mathbf{h}$ by the respective operators $\hat{\mathbf{S}},\hat{\mathbf{I}},\hat{\mathbf{h}}$ henceforth. In the following section, we will consider an arbitrary anisotropy, which allows us to draw conclusions for a graphene QD subjected to either a perpendicular ($\perp$) or a parallel ($\parallel$) oriented external magnetic field.

In any case, the respective field gives rise to a large Zeeman-splitting $b_{S}=\hbar\gamma_{S}B_{z}$ of the electron states, which in turn forbids real spin flip processes due to energy conservation. As a consequence of this, the Hamiltonian can be split in an unperturbed part $\oH_{0}$ consisting of all Zeeman terms and a perturbative part $\oH_{V}$ containing the virtual flip-flop processes of the HI.

If we choose the quantization axis along the magnetic field, the unperturbed Hamiltonian reads:
\begin{equation}
\oH_{0}
=
b_{I}\hat{I}_{z} + b_{S}\hat{S}_{z} + \lz \hat{h}_{z} \oS_{z}
\,\text{,}
\label{eq:H0}
\end{equation}
where the total z-component of the nuclear spins is given by $\hat{I}_{z}=\sum_{k}\hat{I}_{k,z}$. Transforming to a rotating reference frame\cite{Coish2004}, one can eliminate the nuclear Zeeman-term $b_{I}\hat{I}_{z}$ in Eq. (\ref{eq:H0}), which allows to write the unperturbed Hamiltonian $\oH_{0}$ in compact form:
\begin{equation}
\oH_{0}
=
\on\oS_{z}
\,\text{,}
\label{eq:H0g}
\end{equation}
where the Zeeman-energy operator is given by
\begin{equation}
\on = b_{S} - b_{I} + \lz\hz
\,.
\label{eq:onpe}
\end{equation}

The perturbative part $\oH_{V}$ of the Hamiltonian can be written as
\begin{align}
H_{V}
& =
\lx \hat{h}_{x}\hat{S}_{x} + \ly \hat{h}_{y}\hat{S}_{y}
\nonumber\\
& =
\frac{1}{2} \Big( \gp\hat{S}_{-} + \gm\hat{S}_{+}  \Big)
\,,
\label{eq:H_V}
\end{align}
where we introduced raising and lowering operators $S_{\pm}=S_{x}\pm\I S_{y}$ and generalized nuclear operators $\gpm$, which are given by
\begin{align}
\gpm
& =
\frac{1}{2} \left[\left(\lx \pm \ly \right) \hp + \left(\lx \mp \ly \right) \hm \right]
\nonumber\\
& \equiv
\frac{1}{2} \left[\lpm \hp + \lmp \hm \right]
\,.
\label{eq:gpm}
\end{align}
Note, that the bare nuclear magnetic field operators $\hat{h}_\pm = \hat{h}_x\pm \I\hat{h}_y$ can also be considered in terms of single nuclear spin raising and lowering operators located at sites $k$: $\hpm=\sum\limits_{k} A_{k} \hat{I}_{k,\pm}$.

From Eq. (\ref{eq:gpm}), it becomes obvious, that there are two classes of anisotropy with $\lx=\ly$ and $\lx\neq\ly$, respectively. In the former case, the generalized operators $\gpm\propto\hpm$ are just multiples of the bare nuclear magnetic field operators, whereas in the later case these operators become linear combinations of both raising and lowering operators $\hpm$, giving rise to more flip-flop processes.

\subsection{Initial condition}
\label{subsec:initcon}

The time evolution of the combined system, consisting of the electron spin and $N$ nuclear spins, is given by the action of the total Hamiltonian $\oH=\oH_{0}+\oH_{V}$ on the initial state of the system at time $t=0$, which is assumed to be a product state of the form:
\begin{equation}
\ket{\psi\left(0\right)}
=
\ket{n}\otimes\ket{\chi}
\,\text{.}
\label{eq:initstate}
\end{equation}
The z-projection of the electron spin is either parallel ($\chi=\;\ua$) or anti-parallel ($\da$) to the magnetic field, while the nuclear spin bath is prepared in a so-called narrowed state at the beginning of an experiment in order to prevent a fast decay of the electron spin\cite{Schliemann2002,Schliemann2003}.
As discussed in detail in Sec. \ref{sec:key_res}, this narrowing can be created by measuring the nuclear spin system into an eigenstate of the z-component of the nuclear magnetic field $h_{z}$:
\begin{equation}
\hz\ket{n}
=
\av{\hz}_{n}\ket{n}
\,\text{.}
\label{eq:inistat_nuc}
\end{equation}
In general, this narrowed state is a superposition of many degenerate eigenstates $\ket{n_j}$ having all the same polarization $p$:
\begin{align}
\ket{n}				& = \sum_{j=1}^{g_n}\alpha_j \ket{n_j},		&
\hat{h}_{z}\ket{n_{j}}=\av{\hz}_{n}\ket{n_{j}}
\text{,}
\label{eq:initstat_nuc_superpos}
\end{align}
The corresponding eigenvalue of the nuclear field operator $\hz$ is given by\cite{Coish2004}:
\begin{equation}
\av{\hz}_{n}
=
p\, I\, n_{I}\, A
\,\text{.}
\label{eq:hznnfin}
\end{equation}
This result is obtained by converting sums into integrals using a continuum limit, which is valid for times $t\ll\sqrt{N/2}\;\tau_{HI}$, where the corrections to the integrals are still small. 
\begin{table}
  \begin{ruledtabular}
    \begin{tabular}{*4{>{$}c<{$}}}
    \\
    \text{Symbol}	& \text{Definition}			&  \text{Symbol}	& \text{Definition}			\\
    \\
    \hline
    \\
    \epsilon_{HI}	& |\lz| n_{I} A\,/\,2N			& \Delta	& n_{I}A\,/\,2\ons				\\ \\
    \tau_{HI}		& 2N\hbar\,/\,|\lz| n_{I} A		& \Dti		& \Delta \cdot(\lx^{2}+\ly^{2})\,/\,2|\lz|	\\ \\
    \cpm 		& [1\mp p]\,/\,2			& \delta	& \Delta^{2}\,/\,N				\\ \\
    \Nti		& N (\lx^{2}+\ly^{2})\,/\,2\lz^{2}	& \dti		& \delta \cdot(\lx^{2}+\ly^{2})\,/\,2 		\\ \\
    \end{tabular}
  \end{ruledtabular}
\caption{Important symbols used in the text. In the main part of this article, all energies (times) are measured in units of $\epsilon_{HI}$ ($\tau_{HI}$), which is the typical energy (time) scale of a single HI process. The parameters $\cpm$ arise from expectation values of the nuclear bath operators $\hpm$ for spin $1/2$ nuclei, which are characterized by their polarization $p$. The effective number $\Nti$ is introduced to rewrite the denominator given in Eq. (\ref{eq:denpm}) in a generalized form for arbitrary anisotropy $(\lx,\ly,\lz)$. The total Zeeman energy is given by $\ons=b_{S}-b_{I}+\lz\av{\hz}_{n}$, where $b_{S}$ ($b_{I}$) is the electron (nuclear) Zeeman energy and $\av{\hz}_{n}=p\, I\, n_{I}\, A$. For large external magnetic fields, the small parameter $\Delta\ll1$ determines the perturbative regime, while the second small parameter $\delta$ quantifies the non-Markovian corrections. The shorthands $\tilde{\Delta}$ and $\tilde{\delta}$ allow a compact notation of our results.}
\label{tab:symbols}
\end{table}
%

In the following, we introduce dimensionless quantities by measuring energies in units of $\epsilon_{HI}=|\lz| n_{I} A/2N$, where $N=n_{I}N_{tot}$ is the actual number of nuclear spins within the dot. This typical energy scale for a single HI process also defines the corresponding time scale $\tau_{HI}=\epsilon_{HI}/\hbar = 2N\hbar/|\lz| n_{I} A$, which serves as a measure for times henceforth. For clarity, we give a summarizing list of the most important symbols in Tab. \ref{tab:symbols}.

\section{Method}
\label{sec:method}

\subsection{Nakajima-Zwanzig-equation}
\label{subsec:Nakajima}

A common tool to investigate systems like the one considered in this article is the density matrix formalism. The time evolution of the total system is described by the von-Neumann equation $\dot{\rt}\left(t\right)=-\I[\oH,\rt\left(t\right)]=-\I\cL\rt\left(t\right)$, where the Liouville-(super-)operator $\cL=\cL_{0}+\cL_{V}$ also splits in a perturbed and an unperturbed part. The initial density matrix is determined by the initial state of our system given in Eq. (\ref{eq:initstate}): $\rt\left(0\right)=\ket{\psi(0)}\bra{\psi(0)}=\ri \otimes \rs\left(0\right)$. In order to find the dynamics of the electron subsystem only, one can rewrite this differential equation in terms of the Nakajima-Zwanzig master equation\cite{Fick1990,Breuer2002}, where a projection super-operator $\cP \oO = \ri \cdot \Tr_{I} \oO$ is used to perform a partial trace over the nuclear subsystem. This choice of the projector preserves both the initial condition $\cP\rt\left(0\right)=\rt\left(0\right)$ and the electron spin expectation values\cite{Coish2004,Fick1990} $\langle S_{\beta}\rangle\left(t\right)=\Tr\left[S_{\beta}{\cP}\rt(t)\right]=\Tr\left[S_{\beta}\rt(t)\right]$, $\beta=z,\pm$. For later convenience, we also define the complement projector $\cQ$ by $\cP+\cQ=\mathds{1}$. Finally, the equation of motion can be written as
\begin{equation}
\dot{\rt}_{S}\left(t\right)
=
-\I \cL_{0}^{n} \rs\left(t\right) - \I \int \limits_{0}^{t} \D t' \; \Sigma_{S}\left(t-t'\right)\rs\left(t'\right)
\text{,}
\label{eq:NZw_rho_s(t)}
\end{equation}
where the new Liouville operator $\cL^{n}_{0} \oO_{S}=[\ons \cdot \hat{S}_{z},\oO_{S}]_{-}$ originates from the action of the Zeeman energy operator $\on$ on the initial nuclear state $\ri$: $\ons=\Tr_{I}\left(\on\cdot\ri\right)$. While this first part describes the precession of the electron spin in the effective magnetic field, the second term describes the flip flop processes arising from the HI. In order to dissolve the convolution between the self-energy $\Sigma_{S}\left(t\right)$ and the spin density matrix, we perform a Laplace transformation to obtain a much simpler algebraic form:
\begin{widetext}
\begin{equation}
s\rs\left(s\right)-\rs\left(0\right)
=
-\I\,\cL_{0}^{n}\rs\left(s\right)-\I\,\Sigma_{S}\left( s\right)\rs\left(s\right)
\text{,}
\label{eq:rho_s(s)}
\end{equation}
\begin{equation}
\Sigma_{S}(s)
=
-\I\,\Tr_I\left[\cL_{V} \left\lbrace \sum \limits^{\infty}_{l=0} \frac{1}{s+\I\cQ \cL_{0}} \left(-\I\cQ \cL_{V} \frac{1}{s+\I\cQ \cL_{0}}\right)^{l}\right\rbrace \cL_{V}\ri\right]
\equiv
\sum \limits^{\infty}_{j=2} \Sigma_{S}^{\left(j\right)}\left( s\right) 
\,\text{.}
\label{eq:Sigma_laplace} 
\end{equation}
\end{widetext}
The series for the self-energy $\Sigma_{S}\left( s\right)$ in powers of the interaction Liouvillian $\cL_{V}$ is obtained by inserting the Laplace-transformed Schwinger-Dyson identity. For the explicit calculation of these self-energy terms, it is convenient to choose a set of 4 basis vectors given by $\hat{u}_{\ua,\da}=\frac{1}{2}\left(\hat{\sigma}_{0}\pm\hat{\sigma}_{z}\right)$ and $\hat{u}_{\pm}=\frac{1}{2}\left(\hat{\sigma}_{x}\pm\hat{\sigma}_{y}\right)$, where $\left\lbrace \hat{\sigma}_{i} \right\rbrace_{i=x,y,z}$ are the $2\times2$-Pauli-matrices and $\sigma_{0}=\mathds{1}$ is the identity matrix. In this basis, the electron spin density matrix forms a 4-component vector
\begin{align}
\rs	& = \rho_\ua \hat{u}_{\ua} + \rho_\da \hat{u}_{\da} + \av{S_{+}}\hat{u}_{-} + \av{S_{-}}\hat{u}_{+}
\nonumber\\
	& = \left( \rho_\ua,\rho_\da ,\av{S_{+}},\av{S_{-}}\right)^{T}
\text{.}
\label{eq:density_matrix_vector}
\end{align}
The unperturbed Liouvillian is given by a diagonal matrix $\cL_{0}=\frac{1}{2}\cdot\mathrm{diag}\left(\cL_{-},-\cL_{-},-\cL_{+},\cL_{+}\right)$, where the new operators
\begin{equation}
\cL_{\pm}\oO_{I}=[\on,\oO_{I}]_{\pm}
\label{eq:cLpm}
\end{equation}
are defined by their action on an arbitrary nuclear spin operator $\oO_{I}$. The perturbative Liouvillian $\cL_{V}$ has a $4\times4$ off-diagonal form containing the generalized nuclear magnetic field operators $\gpm$. For further details of this calculation, we refer to Appendix A of Ref. \onlinecite{Coish2004} and note that the bare nuclear magnetic field operators $\hpm$ there correspond to the generalized operators $\gpm$ of our model. In general, all contributions $\Sigma^{\left(2j+1\right)}\left( s\right)$ to the self-energy containing odd powers of $\cL_{V}$ vanish , since only virtual flip-flop processes are allowed by energy-conservation. Each even summand $\Sigma^{\left(2j+2\right)}\left( s\right)$ is associated with a small parameter\cite{Coish2004} $\Delta^{j}$, where $\Delta$ is given by:
\begin{equation}
\Delta
=
n_{I}A\,/\,2\ons
\,,
\label{eq:Delta}
\end{equation}
Note, that one finds $\Delta\propto N/\ons$, if one measures the Zeeman energy of the effective field in units of $\epsilon_{HI}$. Experimentally, this parameter is related to the ratio of the nuclear and the external magnetic fields $A$ and $b_{S}$, respectively, giving rise to a small $\Delta\ll1$ for large external fields.  

In this parameter regime, all orders higher than second order are strongly suppressed and, thus, can be neglected. As we show below, neglecting terms of order $O(\Delta)$ limits the range of validity of our analysis to times of order $t\ll\Delta^{-1}\tau_{HI}$. In order to extent to longer times, it would be necessary to include\cite{Coish2010} higher orders of the self-energy, which is however beyond the scope of this article. 

The self-energy exhibits to all orders $\cL_{V}$ a $4\times4$ structure
\begin{equation}
\Sigma(s)
=
\left(
  \begin{array}{cccc}
  \Sigma_{\ua\ua}(s)	& \Sigma_{\ua\da}(s)	& 0 			& 0  		\\
  \Sigma_{\da\ua}(s)	& \Sigma_{\da\da}(s)	& 0 			& 0 		\\
  0 			& 0 			& \Sigma_{++}(s)	& \Sigma_{+-}(s)\\
  0 			& 0 			& \Sigma_{-+}(s) 	& \Sigma_{--}(s)
  \end{array}
\right)
\,\text{,}
\label{eq:self_E4x4}
\end{equation}
which shows a block-diagonal form indicating, that the longitudinal and transverse subspaces are decoupled. In second order, the self-energy takes the specific form:
\begin{align}
\Sigma^{(2)}_{\ua\ua}(s)	& = -\frac{\I}{4}\Tr_I\left(\gm\Fua\gp\ri + \gm\gp\Fda\ri \right)
\,,
\label{eq:Sigma_uu}
\\
\Sigma^{(2)}_{\ua\da}(s)	& = \phantom{-}\frac{\I}{4}\Tr_I\left(\gp\Fda\gm\ri + \gp\gm\Fua\ri\right)
\,,
\label{eq:Sigma_ud}
\\
\Sigma^{(2)}_{\da\da}(s)	& = -\Sigma^{(2)}_{\ua\da}(s)	\qquad\Sigma^{(2)}_{\da\ua}(s) = - \Sigma^{(2)}_{\ua\ua}(s)
\,,
\label{eq:Sigma_dd_du}
\\
\Sigma^{(2)}_{++}(s)		& = -\frac{\I}{4}\Tr_I\left( \gp\Gua\gm\ri + \gm\gp\Gda\ri \right)
\,,
\label{eq:Sigma_pp}
\\
\Sigma^{(2)}_{+-}(s)		& = \phantom{-}\frac{\I}{4}\Tr_I\left( \gp\Gda\gp\ri + \gp\gp\Gua\ri \right)
\,,
\label{eq:Sigma_pm}
\\
\Sigma^{(2)}_{-+}(s)		& = \phantom{-}\frac{\I}{4}\Tr_I\left(\gm\Gua\gm\ri + \gm\gm\Gda\ri \right)
\,,
\label{eq:Sigma_mp}
\\
\Sigma^{(2)}_{--}(s)		& = -\frac{\I}{4}\Tr_I\left(\gm\Gda\gp\ri + \gp\gm\Gua\ri\right)
\label{eq:Sigma_mm}
\,\text{.}
\end{align}
The super-operators of the electron spin dynamics are given by
\begin{equation}
\Fud\left(\cL_{+}\right)=\left(s - \I\alpha_{\ua,\da}\cQ \cL_{+}/2\right)^{-1}
\label{eq:Fud}
\end{equation}
and
\begin{equation}
\Gud\left(\cL_{-}\right)=\left(s + \I\alpha_{\ua,\da}\cQ \cL_{-}/2\right)^{-1}
\,,
\label{eq:Gud}
\end{equation}
where $\alpha_{\ua,\da}=\pm1$. Using these equations in combination with Eqs. (\ref{eq:gpm}) and (\ref{eq:cLpm}), we can compute all parts of the self-energy for arbitrary anisotropy, which is presented in detail in the Appendix. The second order self-energy is given by:
\begin{align}
\Sigma_{\ua\ua}^{(2)}(s)	
& =
-\I N \frac{\lm^{2}}{4\,\lz^{2}} \cm \left[I_{-\tlz}\left(s-\I\ons\right) + I_{+\tlz}\left(s+\I\ons\right)\right]
\nonumber\\
& \phantom{=}\,
-\I N \frac{\lp^{2}}{4\,\lz^{2}} \cp\left[ I_{+\tlz}\left(s-\I\ons\right) + I_{-\tlz}\left(s+\I\ons\right)\right] 
\,,
\label{eq:Sigma_uu_c}
\\
\Sigma_{\ua\da}^{(2)}(s)	
& =
\phantom{-} \I N \frac{\lm^{2}}{4\,\lz^{2}} \cp\left[ I_{+\tlz}\left( s-\I\ons\right) + I_{-\tlz}\left(s+\I\ons\right)\right] 
\nonumber\\
& \phantom{=}\,
+\I N \frac{\lp^{2}}{4\,\lz^{2}} \cm\left[ I_{-\tlz}\left( s-\I\ons\right) + I_{+\tlz}\left(s+\I\ons\right) \right] 
\,,
\label{eq:Sigma_ud_c}
\\
\Sigma_{\da\da}^{{(2)}}(s)
& =
-\Sigma_{\ua\da}^{(2)}(s),\qquad \Sigma_{\da\ua}^{(2)}(s) = - \Sigma_{\ua\ua}^{(2)}(s)
\,,
\label{eq:Sigma_dd_du_c}
\end{align}
\begin{align}
\Sigma_{++}^{(2)}(s)
& =
-\I N \frac{\lm^{2}+\lp^{2}}{4\,\lz^{2}} \left[\cp I_{-\tlz}\left(s\right) + \cm I_{+\tlz}\left(s\right)\right] 
\,,
\label{eq:Sigma_pp_c}
\\
\Sigma_{--}^{(2)}(s)
& =
-\I N \frac{\lm^{2}+\lp^{2}}{4\,\lz^{2}} \left[\cm I_{-\tlz}\left(s\right) + \cp I_{+\tlz}\left(s\right)\right]
\,,
\label{eq:Sigma_mm_c}
\end{align}
\begin{align}
\Sigma_{+-}^{(2)}(s)
& =
\I N \frac{\lm\lp}{2\lz^{2}} \left[\cm I_{-\tlz}\left(s\right) + \cp I_{+\tlz}\left(s\right)\right]=\nonumber\\
& =
-\frac{2\lm\lp}{\lm^{2}+\lp^{2}}\Sigma_{--}^{(2)}(s)
\,,
\label{eq:Sigma_pm_c}
\\
\Sigma_{-+}^{(2)}(s)
& =
\I N \frac{\lm\lp}{2\,\lz^{2}} \left[\cp I_{-\tlz}\left(s\right) + \cm I_{+\tlz}\left(s\right)\right]
\nonumber\\
& =
-\frac{2\lm\lp}{\lm^{2}+\lp^{2}}\Sigma_{++}^{(2)}(s)
\text{,}
\label{eq:Sigma_mp_c}
\end{align}
where the coefficients $\cpm$ arise from the expectation values of the operators $\hpm$ with respect to nuclear magnetic field eigenstates\cite{Barnes2011}. The parameter $\tlz=\mathrm{sign}(\lz)$ takes into account the effect of the sign of the longitudinal anisotropy coefficient $\lz$. In the following all calculations are executed using $\tlz=1$, since the results for $\tlz=-1$ can be easily derived from these computations as we show at the end of the following section.

For a uniformly polarized spin $1/2$ system like graphene, the coefficients $\cpm$ take the simple form $\cpm=[1\mp p]/2$ depending on the polarization $p$. Moreover, the computation of the self energy parts creates super-operator matrix elements $[\mathcal{G}_{j}]_{qn}$ with respect to nuclear eigenstates $\ket{q}$,$\ket{n}$, which in turn give rise to the functions
\begin{equation}
I_{\pm}\left(s\right)=s\left[\log\left(s\mp i\right)-\log\left(s\right)\right]\pm i
\,\text{.}
\label{eq:I(s)continous}
\end{equation}
These functions are calculated applying the same continuum limit used above to obtain Eq. (\ref{eq:hznnfin}).

Analyzing the above Eqs. (\ref{eq:Sigma_uu_c}) to (\ref{eq:Sigma_mp_c}) in more detail, one finds that all longitudinal self-energy parts have always finite values, whereas the off-diagonal transverse parts $\Sigma_{\pm\mp}(s)$ vanish for $\lx=\ly$. In contrast to this, all transverse parts contribute for $\lx\neq\ly$, where the anisotropy of the HI does not only change the prefactors, but also gives rise to additional contributions to the transverse spin component $\av{S_{+}}(t)$.

\subsection{Inverse Laplace transform}
\label{subsec:invlapla}

\begin{table}
  \begin{ruledtabular}
    \begin{tabular}{*{5}{>{$}c<{$}}}
    \\
    \text{Symbol}	& \perp			& \parallel 					& (\lx,\ly,\lz)
    \\
    \\
    \hline
    \\
    \\
    \av{S_{+,1}}_{0}	& \av{S_{+}}_{0}	& \av{S_{+}}_{0}-\frac{3}{5}\av{S_{-}}_{0}	& \av{S_{+}}_{0} -\frac{\lx^{2}-\ly^{2}}{\lx^{2}+\ly^{2}}\av{S_{-}}_{0}
    \\
    \\
    \av{S_{+,2}}_{0}	& 0			& \frac{3}{5}\av{S_{-}}_{0}			& \frac{\lx^{2}-\ly^{2}}{\lx^{2}+\ly^{2}}\av{S_{-}}_{0}
    \\
    \\
    \av{S_{z,1}}_{0}	& \av{S_{z}}_{0}	& \av{S_{z}}_{0}				& \av{S_{z}}_{0}
    \\
    \\
    \av{S_{z,2}}_{0}	& 1			& \frac{4}{5}					& \frac{2\lx\ly}{\lx^{2}+\ly^{2}}
    \\
    \\
    \end{tabular}
  \end{ruledtabular}
\caption{Prefactors of the different models for graphene in perpendicular and parallel magnetic field and for a general anisotropy $\lbrace \lambda_{i}\rbrace$. Note that there is no special choice of the initial values $\av{S_{\beta,\lambda}}_{0}$ of the electron spin.}
\label{tab:num_pref}
\end{table}

With the specific form of the self-energy in second order, we are able to calculate the time evolution of the electron spin density matrix by solving Eq. (\ref{eq:rho_s(s)}). The expectation values for the transverse and longitudinal spin components in Laplace space are obtained by
\begin{equation}
\av{\oS_{\beta}}(s)=\Tr_{S}\Big(\oS_{\beta}\rs(s)\Big)
\label{eq:exp_vals}
\,,
\end{equation}
where $\beta=z,\pm$. The transverse components generally consist of two parts
\begin{equation}
\av{\oS_{\pm}}\left(s\right)
=
\frac{\av{S_{\pm}}_{0}}{D_{\pm}\left(s\right)}+\frac{\I\Sigma^{(2)}_{\pm\mp}\left(s\right)\av{S_{\mp}}_{0}}{D_{\pm}\left(s\right)D_{\mp}\left(s\right)}
\,\text{,}
\label{eq:Spm(s)}
\end{equation}
where the denominator functions are given by
\begin{equation}
D_\pm\left(s\right)=s{\mp}\I\omega_n + \I\Sigma^{(2)}_{\pm\pm}\left(s\right)
\,\text{.}
\label{eq:dpm}
\end{equation}
With the introduction of the effective nuclear number $\Nti=N(\lx^{2}+\ly^{2})/2\lz^{2}$, these functions are formally equal for arbitrary anisotropy. Hence, the mathematics are greatly simplified by the use of these general denominators given in Eq. (\ref{eq:denpm}) below. Using Eqs. (\ref{eq:Sigma_pm_c}) and (\ref{eq:Sigma_mp_c}), one can rewrite the transverse spin component in terms of two generalized functions:
\begin{align}
\av{\oS_{\pm}}\left(s\right)
& =
\av{S_{\pm,1}}\left(s\right) + \av{S_{\pm,2}}\left(s\right)
\nonumber\\
& =
\frac{\av{S_{\pm,1}}_{0}}{D_{\pm}\left(s\right)}
+
\frac{\av{S_{\pm,2}}_{0}\cdot \left(s\pm \I\ons\right)}{D_{\pm}\left(s\right)D_{\mp}\left(s\right)}
\,\text{,}
\label{eq:Spm_gen}
\end{align}
where the prefactors for different cases are listed in Tab. \ref{tab:num_pref}.

In the same manner, we will reformulate the longitudinal spin component
\begin{equation}
\av{\oS_{z}}\left(s\right)
=
\frac{\av{S_{z}}_{0}}{D_{z}\left(s\right)} + \frac{N_{z}\left(s\right)}{D_{z}\left(s\right)}
\,\text{,}
\label{eq:Sz(s)}
\end{equation}
where the denominator function reads
\begin{equation}
D_{z}\left(s\right)
=
s + \I\Big[\Sigma^{(2)}_{\ua\ua}\left(s\right) - \Sigma^{(2)}_{\ua\da}\left(s\right)\Big]
\label{eq:dz}
\end{equation}
and the numerator of the second part is given by
\begin{equation}
N_{z}\left(s\right)
=
-\frac{\I}{2\,s} \left[\Sigma^{(2)}_{\ua\ua}\left(s\right) +  \Sigma^{(2)}_{\ua\da}\left(s\right)\right]
\,\text{.}
\end{equation}
Again, we define two more functions
\begin{align}
\av{\oS_{z}}\left(s\right)
& =
\av{S_{z,1}}\left(s\right) + \av{S_{z,2}}\left(s\right)
\nonumber\\
& =
\frac{\av{S_{z,1}}_{0}}{D_{z}\left(s\right)} + \av{S_{z,2}}_{0} \frac{N_{z}\left(s\right)}{D_{z}\left(s\right)}
\,\text{.}
\label{eq:Sz_gen}
\end{align}
By extending the functional dependencies of all generalized functions given in Eqs. (\ref{eq:Spm_gen}) and (\ref{eq:Sz_gen}) by $\ons$ and $\tlz$, one can easily show, that the following relation holds: 
\begin{equation}
\av{S_{\beta,i}}(s,\ons,-\tlz)=\av{S_{\beta,i}}(s,-\ons,\tlz)
\,,
\label{eq:Spmz_rel}
\end{equation}
where $\beta=\pm,z$. As a consequence of this, a negative coupling $\lz<0$ in the longitudinal direction can be effectively regarded as an inversion of the external magnetic field. 

Moreover, further simplifications are possible by relating the longitudinal denominator $D_{z}(s)$ to the denominator $D_{+}(s)$ of the transverse spin parts through a shift of its variable from $s$ to $\tilde{s}=s+\I\,\ons$:
\begin{align}
&
D_{z}\left(\tilde{s}-\I\ons\right)
=
D_{+}\left(\tilde{s}\right) -\I\frac{N}{\tilde{s}-\I2\ons} + O\left(\delta\right)
\,\text{.}
\label{eq:dz2}
\end{align}
As we will show below, the relevant part of the complex plane is given by $|\mathrm{Im}(s)|\leq\ons$. Thus, the second contribution on the right-hand side is nearly a constant of order $O(\Delta)$ for all relevant parts of the complex plane. From Eq. (\ref{eq:denpm}) below, it is clear, that this constant is small only compared to the denominator $D_{+}\left(\tilde{s}\right)$ for $|\mathrm{Im}(\tilde{s})|\leq\ons$. Thus, this correction to $D_{+}\left(\tilde{s}\right)$ will lead to only a slight and, hence, irrelevant shift in the following calculations, which we will neglect for simplicity. However, reaching $\tilde{s}\approx\I\ons$, this approximation, becomes worse, and, hence, we will take further efforts to investigate this regime. In order to be consistent, we have to shift the numerator $N_{z}(s)$ in the same way as the denominator. The shifted numerator and the transverse denominator are explicitly given by
\begin{widetext}
\begin{equation}
D_\pm\left(s\right)
=
s{\mp}\I\omega_n \pm \I\tilde{N}\left(\cm-\cp\right) + \tilde{N}s \cdot \Bigg[\cmp\Big\lbrace \ln\left(s-\I\right) - \ln\left(s\right)  \Big\rbrace + \cpm\Big\lbrace \ln\left(s+\I\right) - \ln\left(s\right)  \Big\rbrace \Bigg]
\,\text{,}
\label{eq:denpm}
\end{equation}
\begin{align}
N_{z}\left(\tilde{s}-\I\ons\right)
& =                                                                                
\frac{\tilde{N}}{2}\frac{1}{\tilde{s}-\I\ons} \Biggl\lbrace \I\left(\cm+\cp \right) + \tilde{s}\bigg[\cm\Big\lbrace \ln\left(\tilde{s}-\I\right) - \ln\left(\tilde{s}\right) \Big\rbrace - \cp\Big\lbrace \ln\left(\tilde{s}+\I\right) - \ln\left(\tilde{s}\right) \Big\rbrace\bigg]\Biggr\rbrace + O\left(\delta\right)
\,\text{.}
\label{eq:nz}
\end{align}
\end{widetext}
The corrections to $N_{z}\left(\tilde{s}-\I\ons\right)$ are of order $O(\delta)$ for $|s|<\ons$ and consequently negligible due to their smallness. In the vicinity of $s\approx \I\ons$, though, we face similar difficulties as for the denominator demanding a careful treatment. 

\begin{figure}
\includegraphics[width=0.25\textwidth]{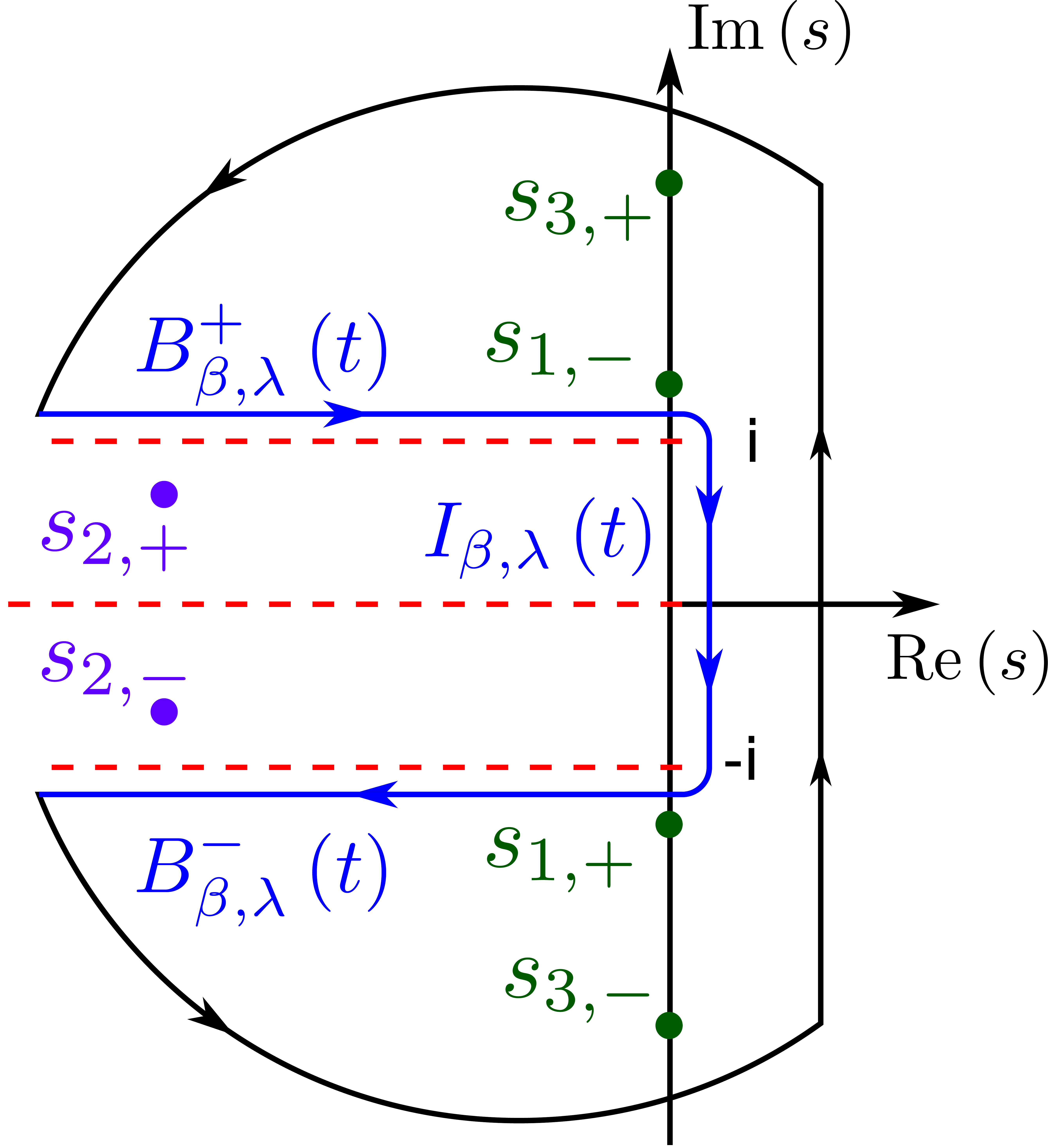}
\caption{(Color online) Illustration of the contour integral in the complex plane with the analytic features of the denominator functions $D_{+}(s)$ and $D_{-}(s)$ consisting of branch-cuts (red dashed lines) and poles (blue and green circles). The integral in Eq. (\ref{eq:S_beta(t)}) is completed to a closed contour by an exponentially vanishing integral over the great circle, integrals  along the upper and lower branch-cuts, $B_{\beta,\lambda}^\alpha\left(t \right)$, and an integral along the imaginary axis $I_{\beta,\lambda}\left(t\right)$. Within this closed contour lie the poles $s_{j,\pm}$, $j=1,3$, whereas the poles $s_{2,\pm}$ are not encircled and, thus, do not contribute to the integral. Note that the poles $s_{j,-}$ are relevant only for the calculation of $\av{S_{2,+}}(t)$.}
\label{fig:contour}
\end{figure}

These peculiarities are especially important for the inverse Laplace transformation, which is applied to recover the time-dependence of the electron spin. This transformation is obtained by evaluating integrals of the form
\begin{equation}
\av{\oS_{\beta,\lambda}}\left(t\right)
=
\frac{\E^{-\I \Omega_{\beta} t}}{2\pi\I}  \int\limits_{\gamma-\I\infty}^{\gamma+\I\infty} \E^{st} \av{\oS_{\beta,\lambda}}\left(s-\I \Omega_{\beta}\right) \D s
\;\text{,}\;
\label{eq:S_beta(t)}
\end{equation}
along the Bromwich contour by means of complex analysis, where the indices are given by $\beta=z,\pm$ and $\lambda=1,2$. The frequency shift
\begin{equation}
\Omega_{\beta}
=
  \begin{cases}
  0 	& \mbox{for}\; \beta = \pm \\
  \ons 	& \mbox{for}\; \beta = z
  \end{cases}
\end{equation}
is relevant only for the $z$ component of the electron spin and is used to simplify the mathematics as explained above. The constant $\gamma\in\mathbb{R}$ is chosen, such that all singularities have a real part smaller than $\gamma$. These singularities are generated by the denominators $D_{\pm}\left(s\right)$ each possessing three zeros at $s_{j,\pm}$ ($j=1,2,3)$, which come in complex-conjugated pairs $s_{j,+}=s^{*}_{j,-}$. Furthermore, there are three branch cuts from $s=0,\pm\I$ to $-\infty$, whose position in the complex plane is illustrated in Fig. \ref{fig:contour}.

In order to evaluate the integral in Eq. (\ref{eq:S_beta(t)}), one can close the contour as depicted in Fig. \ref{fig:contour}, where the integral over the great circle vanishes according to Jordan's lemma. Note that the poles with finite real part $s_{2,\pm}$ are outside of this contour and, hence, do not contribute to the integral. Therefore, the solution of Eq. (\ref{eq:S_beta(t)}) generally consists of residues arising from the remaining poles, $s_{1,\pm}$ and $s_{3,\pm}$, the integrals $B_{\beta,\lambda}^\alpha\left(t\right)$ along the upper and lower branch-cut as well as an integral $I_{\beta,\lambda}\left(t\right)$ along the imaginary axis:
\begin{align}
\av{\oS_{\beta,\lambda}}\left(t\right)
& =
\sum_{s=\lbrace s_{1,\pm},s_{3,\pm}\rbrace} \Res\left[\E^{st}\cdot\av{\oS_{\beta,\lambda}}\left(s\right)\right]
\nonumber\\
& \phantom{=}
-\frac{1}{2\pi \I} \E^{-\I \Omega_{\beta} t} \underbrace{\Big\lbrace\sum_{\alpha}B_{\beta,\lambda}^\alpha\left(t\right) + I_{\beta,\lambda}\left(t\right) \Big\rbrace}_{=P_{\beta,\lambda}(t)}
\,,
\label{eq:S_contour}
\end{align}
where the integrals are explicitly given by
\begin{align}
B_{\beta,\lambda}^\alpha\left(t \right) 
=
\lim_{\eta\rightarrow 0} \E^{\I\alpha t} \int \limits_{-\infty}^{0} \D x\; \E^{xt}
\av{\oS_{\beta,\lambda}}\big(x+\I\alpha(1+\eta)-\I \Omega_{\beta}\big)
\label{eq:B+-}
\end{align}
with $\alpha=\pm1$ and
\begin{equation}
I_{\beta,\lambda}\left(t\right)
=
\I \int \limits_{-1}^{1} \D y\; \E^{\I yt} \cdot\;
\av{\oS_{\beta,\lambda}}\big(\I y -\I \Omega_{\beta}\big)
\,.
\label{eq:I}
\end{equation}
In order to simplify the notation of the results and to make their interpretation easier, we use the short hands $\Nti$, $\Delta$, $\Dti=\Delta \cdot(\lx^{2}+\ly^{2})\,/\,2\lz^{2}$,  $\delta$, and $\dti=\delta \cdot(\lx^{2}+\ly^{2})\,/\,2$ listed in Table \ref{tab:symbols} as well as the relations $\cm+\cp=1$ and $\cm-\cp=p$, which are fulfilled for nuclei with spin $I=1/2$ as considered here.

\section{Results}
\label{sec:results}

\subsection{Inverse Laplace transformation of the first transverse spin part}

\begin{table*}
  \begin{ruledtabular}
    \begin{tabular}{@{\extracolsep{5mm}}*{4}{>{$}l<{$}}}
    \\
      &
      D_{+}\left(s_{i}\right)\stackrel{!}{=}0
      &
      \text{Res}\left[\E^{st}\cdot\av{S_{+,1}}\left(s\right)\right]_{s=s_{i}}
      &
      \text{Res}\left[\E^{\tilde{s}t}\cdot\av{S_{z,2}}\left(\tilde{s}-\I\ons\right)\right]_{\tilde{s}=s_{i}}
    \\
    \\
    \hline
    \\
    \\
      s_{1,+}:
      &
      - \I\left[ 1+2^{-\frac{\cm}{\cp}}\E^{-\frac{1}{\Dti\,\cp}}\right] 
      &
      \av{S_{+,1}}_{0} \left[\Nti\cp  2^{\frac{\cm}{\cp}}\right]^{-1} \E^{-\frac{1}{\Dti\,\cp}- \I t}
      &
      -\frac{\av{S_{z,2}}_{0}}{2} \left[\Nti\cp 2^{\frac{\cm}{\cp}}\right]^{-1} \E^{-\frac{1}{\Dti\,\cp}-\I\left(\ons+1 \right)t}
      \\
    \\
      s_{3,+}:
      &
      \I\ons
      &
      \av{S_{+,1}}_{0} \left[1+\frac{1}{2}\dti\right]^{-1}  \E^{\I\ons t} 
      &
      \\
    \\
    \end{tabular}
  \end{ruledtabular}
\caption{Zeros of the denominator $D_{+}(s)$ and the corresponding residues for the transverse and longitudinal electron spin components $\av{S_{+,1}}(t)$ and $\av{S_{z,2}}\left(t\right)$, respectively. The purely imaginary pole $s_{3,+}$ gives rise to an undamped oscillation around the effective magnetic-field $\ons$. The special case of $\tilde{s}=0$, which corresponds to $s=s_{3,+}=\I\ons$, gives rise to a longtime average ${\av{S_{z}}}_{\infty}$, which we discuss in more detail in the text.}
\label{tab:poles_residues}
\end{table*}

We begin the inverse transformation into the time domain with the first part of the transverse spin component $\av{S_{+,1}}\left(s\right)$,  which is exemplary for the calculation of all other expectation values $\av{S_{\beta,\lambda}}(t)$.

First, we analyze the residues arising from the poles $s_{j,+}$, which are summarized in Tab. \ref{tab:poles_residues}. The pole $s_{3,+}$ located on the imaginary axis gives rise to a purely oscillating term, where the frequency is given by the effective magnetic field $\ons$. This oscillating part corresponds to a simple precession of the electron spin around this magnetic field. 

The calculation of this residue also nicely illustrates in which sense the disregard of the fourth order contribution $\Sigma^{(4)}(s)\propto\Delta$ of the self-energy sets an upper time limit. For simplicity, we assume that this contribution can be described by a complex valued constant $\Sigma^{(4)}(s)\approx\xi\Delta$ of order $O(\Delta)$, where we neglect any dependence on $s$. According to Eq. (\ref{eq:dpm}), this constant shift can be formally treated as a modification of the effective magnetic field $\I\ons\rightarrow\I\ons-\xi\Delta$ giving rise to an additional exponential factor $\exp(-\xi\Delta\cdot t)$. Hence, our predictions are valid only for times $t\ll\Delta^{-1}\tau_{HI}$, for which this factor is irrelevant. In order to extend this limit, one would have to take the fourth order contribution with its full $s$-dependence into account.

This line of arguing is, however, not directly applicable for the oscillating portion originating from the pole $s_{1,+}$, since this pole has a more complicated structure. It is located near the lower branching point at $-\I$, as is illustrated in Fig. \ref{fig:contour}. This residue generates an amplitude, which is exponentially small $\exp[-(\cp\Delta)^{-1}]\ll1$ in a large magnetic field. As a consequence of its smallness, we will neglect this contribution in the following for simplicity. Finally, the pole $s_{2,+}$ does not contribute, because it is outside of the contour.

\begin{figure}
\includegraphics[width=0.5\textwidth]{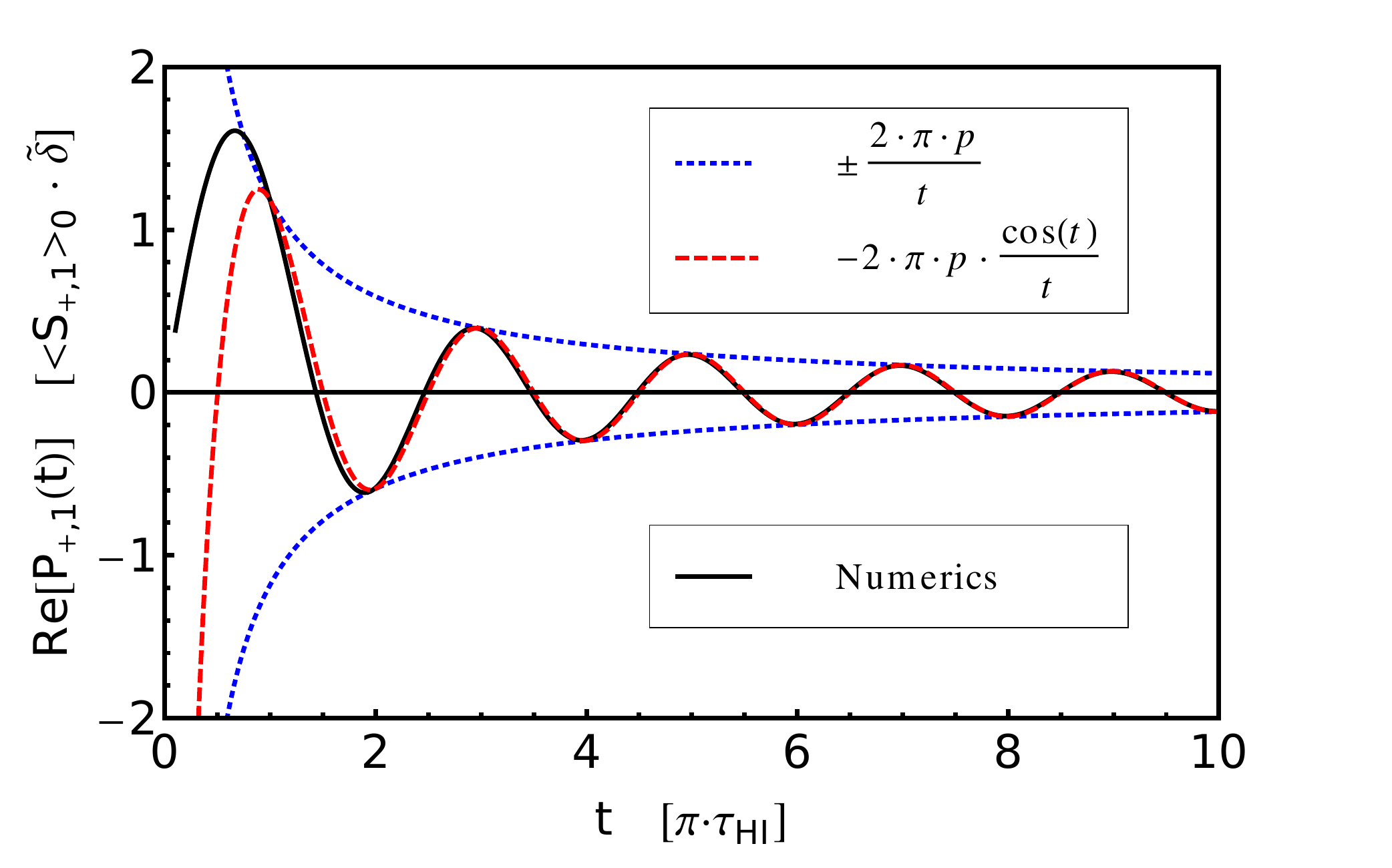}
\caption{(Color online) Real part of $P_{+,1}(t)$ (solid, black) as a function of time obtained by numerical integration. For times $t\gtrsim\tau_{HI}$, it is asymptotically described by an oscillating function (dashed, red), whose amplitude decays with $\sim t^{-1}$ (dotted, blue), where the amplitude is proportional to the polarization $p$. The oscillations show a frequency of $f=\tau_{HI}^{-1}\propto |\lz|A$ determined by the HI.}
\label{fig:S+1_numint_Re}
\end{figure}
\begin{figure}
\includegraphics[width=0.5\textwidth]{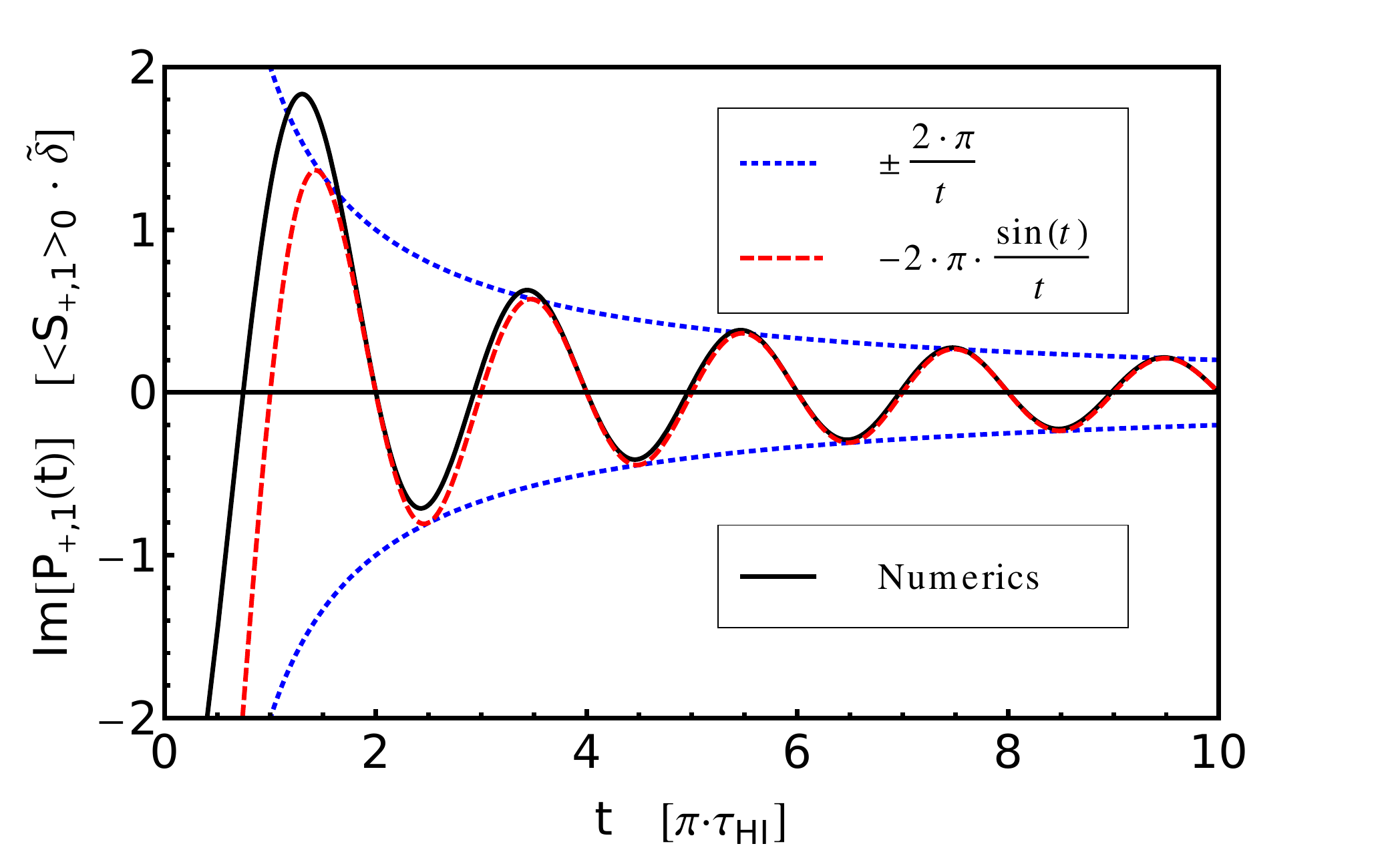}
\caption{(Color online) Imaginary part of $P_{+,1}(t)$ (solid, black) showing a similar behavior as the real part depicted in Fig. \ref{fig:S+1_numint_Re}. In contrast to the real part, the amplitude of the imaginary part does not depend on the polarization.}
\label{fig:S+1_numint_Im}
\end{figure}

Since the integral over the great circle vanishes due to Jordan's lemma, the only remaining, unknown expressions arise from the integrals along the branch-cuts and along the imaginary axis. The calculation of these integrals is, however, mathematically very challenging due to the fact that three different scaling behaviors are involved. The inverse Laplace transformation itself gives rise to an exponential factor $\exp(st)$, while the denominators contain both logarithmic and power law terms hampering analytical solutions to these integrals. Nevertheless, analytical considerations give valuable insights to the structure of the results. First, we checked, that there are no contributions present in $P_{\beta,\lambda}(t)$, which diverge for longer times and, thus, would lead to unphysical results. Moreover, one can testify, that the leading order contributions arising from the branch cut integrals $B^{\beta,\lambda}_{\pm}\left(t \right)$ and from the imaginary integrals $I_{\beta,\lambda}\left(t\right)$ cancel each other leaving terms of order $\delta=\Delta^2/N$. While it is easy to show that the contributions of this order stemming from the branch cut integrals are oscillating with a frequency $\tau_{HI}^{-1}$ determined by the HI, this is not evident for the imaginary integral. Furthermore, there is no obvious way to analytically extract more information on the time dependence of the amplitudes such as the form of a possible decay. Hence, we use numerical methods to find the results for all integrals, which are subsequently summed up in order to give the functions $P_{\beta,\lambda}(t)$ defined in Eq. (\ref{eq:S_contour}) above.

The function being relevant for $\av{S_{+,1}}(t)$ is given by $P_{+,1}(t)$, whose real and imaginary part is plotted in Figs. \ref{fig:S+1_numint_Re} and \ref{fig:S+1_numint_Im}, respectively. For times $t\gtrsim\tau_{HI}$, we find that the sum of the branch-cut contributions is asymptotically described by an oscillating term, whose amplitude is decaying with a power law:
\begin{equation}
\frac{P_{+,1}(t)}{\av{S_{+,1}}_{0}}
=
-2 \pi \I\, \dti\, \Big[\frac{\sin(t)}{t}-\I p \frac{\cos(t)}{t} \Big]
\,.
\label{eq:P+1_final}
\end{equation}

The final result for $\av{S_{+,1}}(t)$ is obtained by summing up the residues and power law contributions according to Eq. (\ref{eq:S_contour}):
\begin{align}
\frac{\av{S_{+,1}}\left(t\right)}{\av{S_{+,1}}_{0}} =
&
\Big[1+(\lx^{2}+\ly^{2})\frac{\delta}{4}\Big]^{-1}  \E^{\I\ons t} +
\nonumber\\
&
\frac{\lx^{2}+\ly^{2}}{2}\;\delta\;\Big[\frac{\sin(t)}{t}-\I p \frac{\cos(t)}{t}\Big]
\,\text{,}
\label{eq:S+1_final}
\end{align}
where we reintroduced the explicit dependence on the anisotropy of the HI using the relations summarized in Tab \ref{tab:symbols}. Note, that a negative coupling $\lz<0$ in $z$-direction can be easily handled using Eq. (\ref{eq:Spmz_rel}).

For the special case of isotropy within the $x$-$y$ plane, Eq. (\ref{eq:Spm_gen}) readily gives $\av{S_{+,2}}=0$. Thus, setting $\lx=\ly$ in the above equation already allows us to interpret the dynamics of this type of systems, which are physically realized in \ga{} or graphene subjected to a perpendicular magnetic field. According to Eq. (\ref{eq:S+1_final}) only a small fraction of order $\delta=\Delta^2/N$ of the initial transverse spin decays and does so in a power law, while most of the transverse spin is preserved and oscillates with a frequency determined by the effective magnetic field $\ons$. Physically this means, that the transverse spin is only little affected by the virtual spin flip-flops, which are strongly suppressed because of the enormous energy difference between the Zeeman splitting and the HI energy.

\subsection{Inverse Laplace transformation of the second transverse spin part}

So far, we have calculated the time-dependence of the first transverse spin part $\av{S_{+,1}}(t)$, which fully describes the behavior of a system with rotational symmetry in the transverse x-y-plane. For a system having a broken rotational symmetry with $\lx\neq\ly$, one has additionally to calculate the second transverse spin function $\av{S_{+,2}}(t)$, which contains two denominator functions $D_{+}(s)$ and $D_{-}(s)$ instead of one. 

Turning first to the residues, we can take advantage of the fact, that the zeros of these denominators come in complex conjugated pairs $s_{j,+}=s^{*}_{j,-}$ because of the relation $D^{*}_{+}(s)=D_{-}(s^{*})$. Therefore, the calculation of the residues is straightforward, resulting in the contributions listed in Tab. \ref{tab:poles_residues_prod}.
\begin{table*}
  \begin{ruledtabular}
    \begin{tabular}{ccc}
    \\
      $D_{\pm}\left(s\right)\stackrel{!}{=}0$
      &
      $\text{Res}\left[\E^{st}\av{S_{+,2}}\left(s\right)\right]_{s=s_{j,+}}$
      &
      $\text{Res}\left[\E^{st}\av{S_{+,2}}\left(s\right)\right]_{s=s_{j,-}=s^{*}_{j,+}}$
    \\
    \hline
    \\
      $s_{1,+}$
      &
      $ \av{S_{2,+}}_{0}\left[\Nti\cp \; 2^{\frac{\cm}{\cp}}\right]^{-1} \;\E^{-\I t}\;\E^{-\frac{1}{\Dti\cp}} $
      &
      $-\av{S_{2,+}}_{0}\left[\Nti\cp \; 2^{\frac{\cm}{\cp}}\right]^{-1}\;\E^{\I t} \;\E^{-\frac{1}{\Dti\cp}} $
      \\
      \\
      $s_{3,+}$
      &
      $\av{S_{+,2}}_{0} \left[1+\frac{1}{2}\dti\right]^{-1} \E^{\I\ons t}$
      &
      $0$
    \\
    \end{tabular}
  \end{ruledtabular}
\caption{Residues of the second transverse spin part $\av{S_{2,+}}(t)$. The poles $s_{j,+}$ are listed in Tab. \ref{tab:poles_residues}. Here the $s_{3,\pm}$-poles produce inequivalent residues due to the rewritten form of $\av{S_{2,+}}\left(s\right)$ in Eq. (\ref{eq:Spm_gen}).}
\label{tab:poles_residues_prod}
\end{table*}
The poles $s_{1,\pm}$ generate exponentially suppressed terms, which are equivalent to the $s_{1,+}$-residue of $\av{S_{+,1}}(t)$. This means in particular, that these residues are also negligible for large magnetic fields. In contrast to these paired terms, the pole $s_{3,+}$ creates a purely oscillating component, while the residue of pole $s_{3,-}$ vanishes identically.

\begin{figure}
\includegraphics[width=0.5\textwidth]{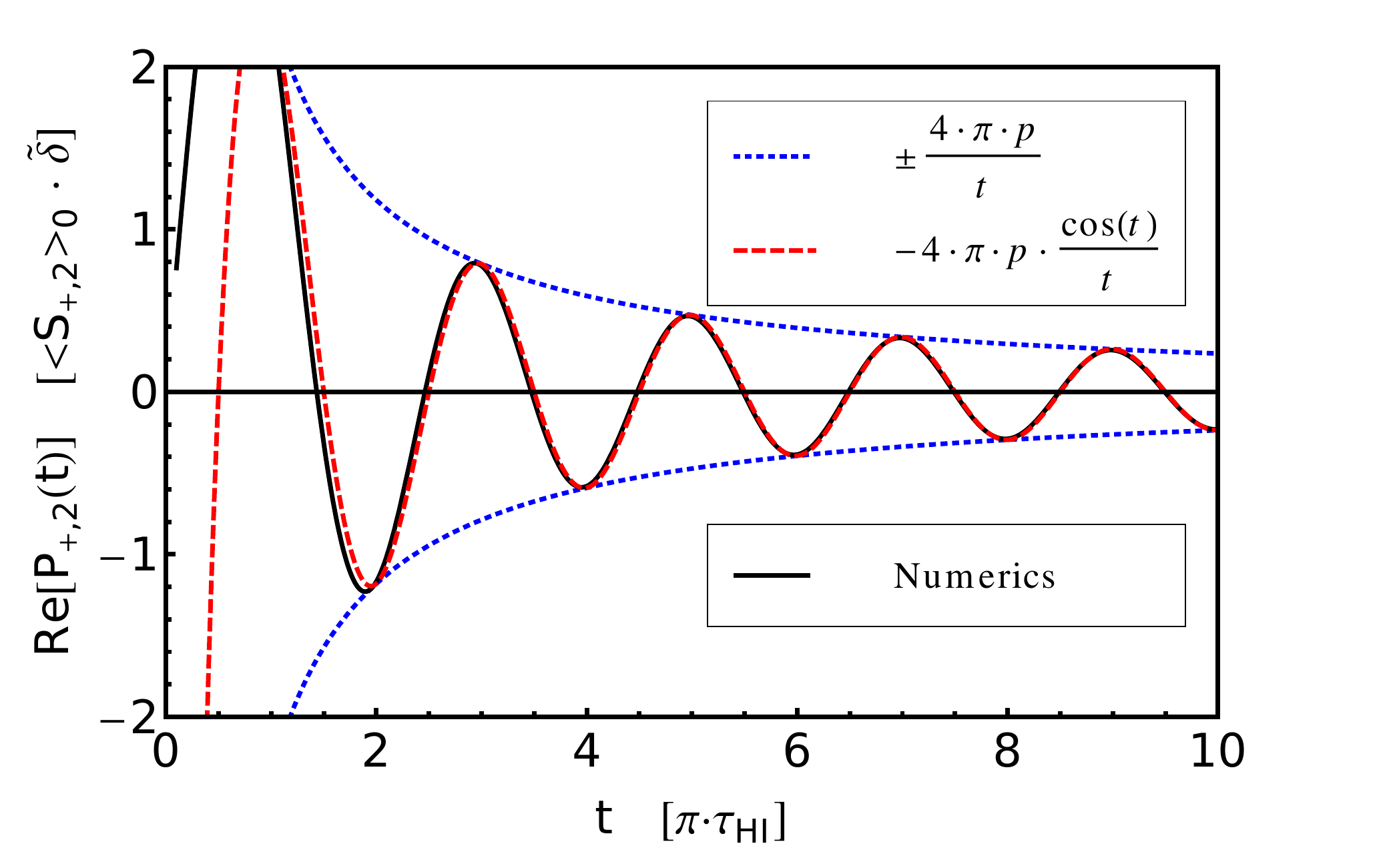}
\caption{(Color online) Real part of $P_{+,2}(t)$ (solid, black) as a function of time obtained by numerical integration. For times $t\gtrsim\tau_{HI}$, it is asymptotically described by an oscillating power law decay (dashed, red; dotted, blue) $\sim t^{-1}$, where the amplitude is proportional to the polarization $p$ and the frequency is given by $f=\tau_{HI}^{-1}$. }
\label{fig:S+2_numint_Re}
\end{figure}

The calculation of the sum of the branch-cut integrals, $P_{+,2}(t)$, is again obtained by numerical integration accompanied by analytical considerations. We find, that only the real part of $P_{+,2}(t)$ has a relevant contribution of order $O(\delta)$, while the imaginary part is much smaller of order $O(\delta/\ons)$ and, thus, neglected. The time evolution of this real part is, up to a factor of two, analogous to the real part of $P_{+,1}(t)$, which can be read off from Fig. \ref{fig:S+2_numint_Re}. 

We find, that the branch-cut contribution can be asymptotically described by 
\begin{equation}
\frac{P_{+,2}(t)}{\av{S_{+,2}}_{0}}
=
-2\pi\I \, \dti \Big[-\I 2 p \frac{\cos(t)}{t} \Big]
\end{equation}
for times $t\gtrsim\tau_{HI}$. In order to obtain the full $\av{S_{+,2}}(t)$ term, we sum up the residues and power law contributions yielding:
\begin{align}
\frac{\av{S_{+,2}}(t)}{\av{S_{+,2}}_{0}} =
&
\Big[1+(\lx^{2}+\ly^{2})\frac{\delta}{4}\Big]^{-1}  \E^{\I\ons t} +
\nonumber\\
&
\frac{\lx^{2}+\ly^{2}}{2}\,\delta \Big[-\I 2 p \frac{\cos(t)}{t} \Big]
\,.
\label{eq:S+2_final}
\end{align}
In combination with the result for $\av{S_{+,1}}(t)$ given in Eq. (\ref{eq:S+1_final}), we are now able to formulate the time dependence of the transverse electron spin component $\av{S_{+}}(t)$ for arbitrary anisotropy and a general initial condition $\av{S_{\pm}}_{0}=\av{S_{x}}_{0}\pm\I\av{S_{y}}_{0}$:
\begin{align}
\av{S_{+}}(t)
& =
\av{S_{+,1}}(t) + \av{S_{+,2}}(t)
\nonumber\\
& =
(\av{S_{x}}_{0}+\I\av{S_{y}}_{0}) \; \Big[1+(\lx^{2}+\ly^{2})\frac{\delta}{4}\Big]^{-1}  \E^{\I\ons t}
\nonumber\\
& +
\;\big(\ly^{2}\av{S_{x}}_{0} +\I \lx^{2}\av{S_{y}}_{0}\big) \; \delta\; \frac{\sin(t)}{t} 
\nonumber\\
& -
\I \big(\lx^{2}\av{S_{x}}_{0} +\I \ly^{2}\av{S_{y}}_{0}\big) \; \delta\; p\; \frac{\cos(t)}{t}
\,.
\label{eq:S+_final}
\end{align}
In the limit $\ly=\lx$, this result reproduces the previous result given in Eq. (\ref{eq:S+1_final}). As in the isotropic limit discussed above, we find that due to the large Zeeman splitting most of the transverse electron spin is preserved and precesses around the effective magnetic field $\ons$. The decaying part, however, differs from the isotropic case and allows us to analyze how the effect of the HI changes as $\lx\neq\ly$. As expected for the transverse expectation value $\av{S_{+}}(t)$, Eq. (\ref{eq:S+_final}) is formally invariant under the exchange $x\leftrightarrow y$, which allows us to discuss the result for a specific choice of $\lx>\ly$ without loss of generality. Modifying the coupling until $\ly/\lx=0$ is reached, the amplitude of the sinusoidal part is more and more dominated by the initial electron spin in $y$ direction, while the amplitude of the cosine term is increasingly governed by the initial component $\av{S_{x}}_{0}$. The reason for this is the broken rotational symmetry in the $x$-$y$-plane due to the anisotropy of the HI. This in turn makes the time evolution of the electron spin dependent on its initial preparation.

Furthermore, the amplitude of the cosine term also depends on the polarization $p$ of the nuclear bath, which quantifies the excess of one nuclear spin orientation over the other. For increasing $p$ one type of the HI-induced scattering processes, for instance $\hp$, becomes more likely, while the other one is suppressed, since its phase space is more and more limited. Presumably, this can explain the polarization dependence of our result. 

However, within the Nakajima-Zwanzig formalism, it is not possible to single out microscopic processes explaining the specific form of the dependence on the anisotropy and the polarization demanding for successive studies using different techniques.

\subsection{Longitudinal electron spin for arbitrary HI}

\begin{figure}
\includegraphics[width=0.5\textwidth]{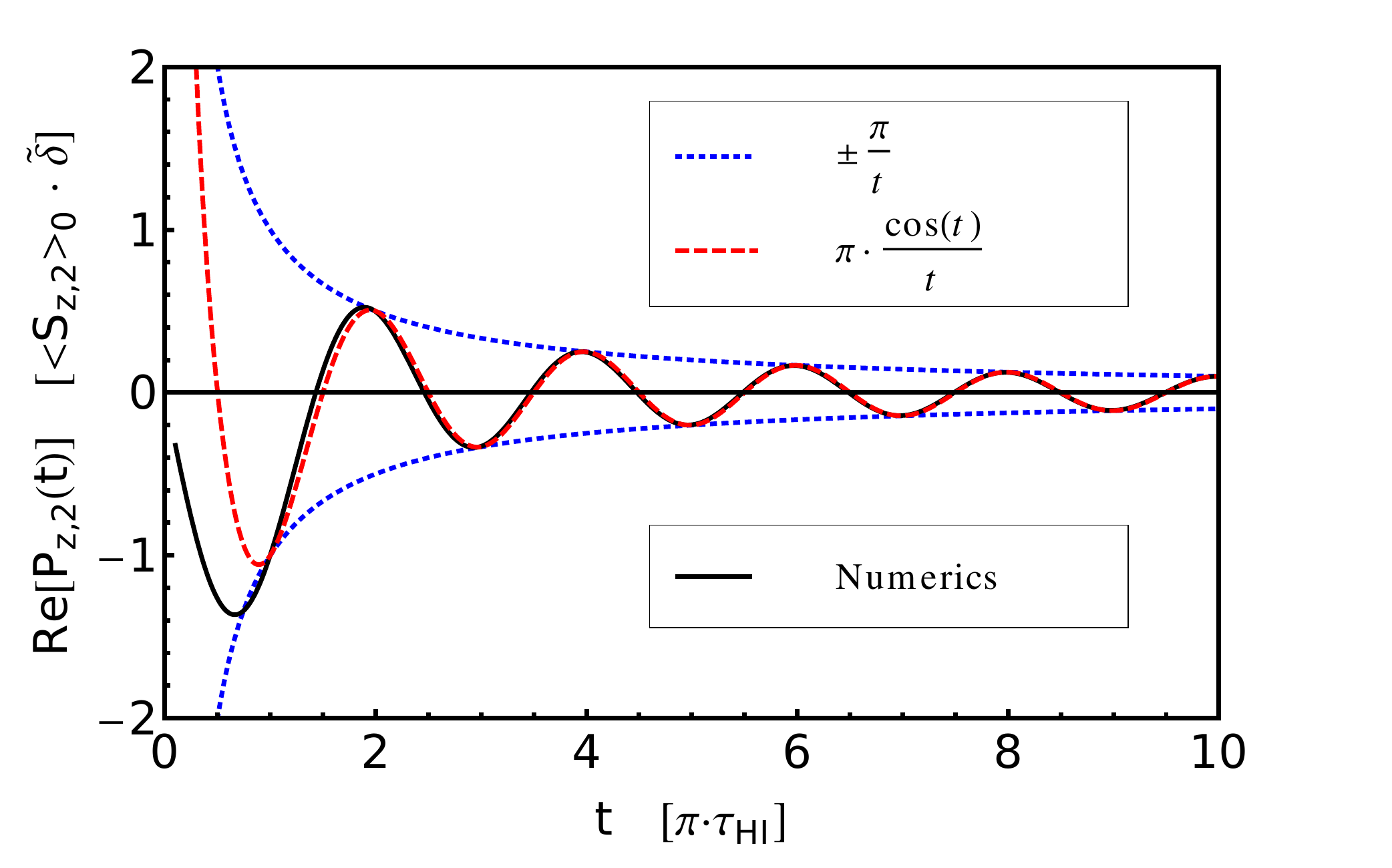}
\caption{(Color online) Real part of $P_{z,2}(t)$ (solid, black) as a function of time obtained by numerical integration showing a similar behavior as $\mathrm{Re}[P_{+,1}(t)]$ depicted in Fig. \ref{fig:S+1_numint_Re}.  In contrast to this, however, its amplitude does not depend on the polarization.}
\label{fig:Sz2_numint_Re}
\end{figure}
\begin{figure}
\includegraphics[width=0.5\textwidth]{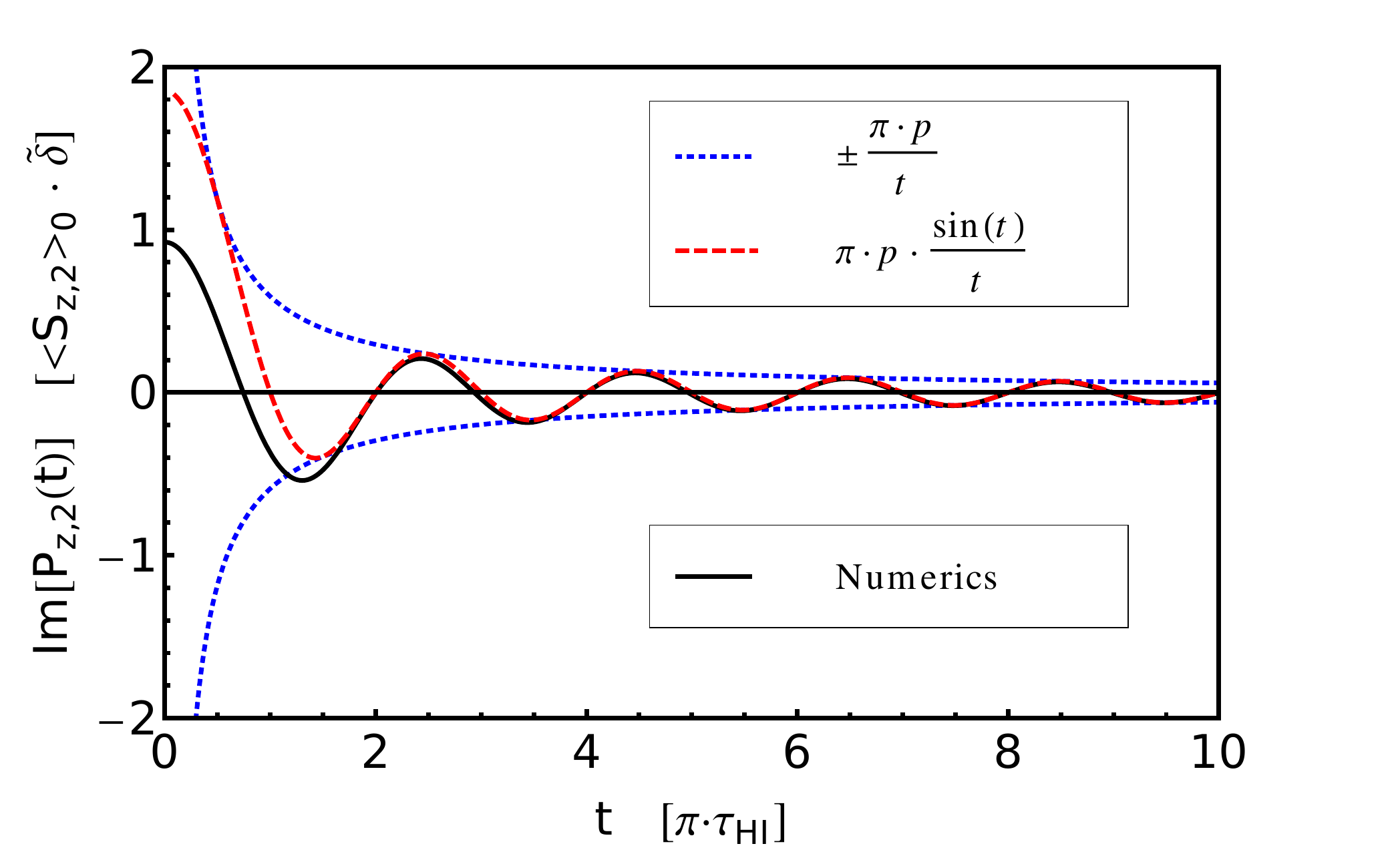}
\caption{(Color online) Imaginary part of $P_{z,2}(t)$ (solid, black) as a function of time obtained by numerical integration showing a similar behavior as $\mathrm{Im}[P_{+,1}(t)]$ depicted in Fig. \ref{fig:S+1_numint_Im}. Deviating from this, the amplitude depends on the polarization of the nuclear bath.}
\label{fig:Sz2_numint_Im}
\end{figure}

Next, we turn to the inverse Laplace transformation of the longitudinal spin components, where we use the shifted denominator and numerator functions given in Eqs. (\ref{eq:dz2}) and (\ref{eq:nz}), respectively, to re-express Eq. (\ref{eq:S_beta(t)}) in terms of the new coordinate $\tilde{s}$:
\begin{align}
\av{S_{z,1/2}}\left(t\right)
& =
\frac{\E^{-\I\ons t}}{2\pi\I} \int\limits_{\gamma-\I\infty}^{\gamma+\I\infty} \E^{\tilde{s}t} \av{S_{z,1/2}}\left(\tilde{s}-\I\ons\right) \D \tilde{s}
\;\text{.}
\end{align}
With this reformulation, the integrand $\av{S_{z,1}}\left(\tilde{s}-\I\ons\right)$ has the same analytic structure like $\av{S_{+,1}}(s)$ for $\mathrm{Im}(\tilde{s})\ll\ons$, and, hence, we can take advantage of previous results. Therefore, the contributions of the poles $s_{1,+}$ as well as the branch-cut and imaginary integrals are readily obtained by replacing the prefactor $\av{S_{1,+}}_{0}$ by $\av{S_{z,1}}_{0}\cdot \exp[-\I\ons t]$ in Tab. \ref{tab:poles_residues} and in Eq. (\ref{eq:P+1_final}). As we discussed extensively in context with the derivation of Eqs. (\ref{eq:dz2}) and (\ref{eq:nz}), the calculation of the residue of the pole $s_{3,+}=\I\ons$ needs a more sophisticated treatment, which will be done in detail below.

In order to find the full time dependence of the longitudinal spin component, we also need to compute the second contribution $\av{S_{z,2}}\left(t\right)$, which follows the same procedure as $\av{S_{+,1}}(t)$. This calculation is in principle easy, but lengthy and is, therefore, not presented in detail. The residue of the pole $s_{1,+}$ is listed in Tab. \ref{tab:poles_residues}, whereas the outcome of the pole at $s_{3,+}=\I\ons$ will be also presented in more detail below. 

The integrals making up $P_{z,2}(t)$ are again calculated numerically. The time dependence of these integrals is presented in Figs. \ref{fig:Sz2_numint_Re} and \ref{fig:Sz2_numint_Im}, from which their power law decaying behavior becomes obvious. Asymptotically this can be described by the following function:
\begin{equation}
\frac{P_{z,2}(t)}{\av{S_{z,2}}_{0}}
=
2\pi\I\;\frac{\dti}{2}\Big[ p\frac{\sin(t)}{t} - \I \frac{\cos(t)}{t} \Big]
\,.
\label{eq:Pz_2}
\end{equation}
Note, that this result is smaller by a factor of two compared to the outcome of $P_{+,1}(t)$ given in Eq. (\ref{eq:P+1_final}) and that the real part depends on the polarization instead of the imaginary part. At last, we have to evaluate the residues at the pole $s_{3,+}$ for both $\av{S_{z,1}}(t)$ and $\av{S_{z,2}}(t)$. If we shift back to the original coordinate system $\tilde{s}\rightarrow s=\tilde{s}-\I\ons$, this residue can be rewritten as a limit for long times according to the properties of the Laplace transform:
\begin{align}
& \phantom{=}\Res_{\tilde{s}=s_{3,+}} \left[\E^{\left(\tilde{s}-\I\ons\right)t}\av{S_{z,1/2}}\left(\tilde{s}-\I\ons\right)\right]
\nonumber\\
& =
\lim_{s\rightarrow0} s\; \E^{st}\av{S_{z,1/2}}\left(s\right)
=
\av{\bar{S}_{z,1/2}}
\,\text{.}
\end{align}
This longtime average of the z-spin component $\bar{\av{S_{z}}}=\av{\bar{S}_{z,1}}+\av{\bar{S}_{z,2}}$ is actually calculated by performing a series expansion in $\left( s\pm \I\ons\right)^{-1}$ of Eq. (\ref{eq:Sz_gen}) and subsequently taking the limit $s\rightarrow0$, which finally leads to:
\begin{align}
\bar{\av{S_{z}}}
& =
\frac{\av{S_{z,1}}_{0}+\frac{\dti\,p}{2}\,\av{S_{z,2}}_{0}+O\left(\frac{N}{\ons^{3}}\right)}{1+\dti + O\left(\frac{N}{\ons^{3}}\right)}
\nonumber\\
& \approx
\frac{\av{S_{z}}_{0}+\lx\ly\;\frac{\delta\,p}{2}}{1+\frac{\lx^2+\ly^2}{2}\delta}
\label{eq:sz_longtime}
\,\text{.}
\end{align}
With this result, all contributions to the longitudinal electron spin component are known: 
\begin{align}
\av{S_{z}}\left(t\right)
& =
\bar{\av{S_{z}}} + \dti \cdot \E^{-\I\ons t} \cdot
\nonumber\\
& \phantom{=}\;
\bigg[\Big\lbrace\av{S_{z,1}}_{0} - \frac{p}{2}\av{S_{z,2}}_{0}\Big\rbrace\, \frac{\sin(t)}{t} +
\nonumber\\
& \phantom{=}\;
-\I\phantom{\bigg[} \Big\lbrace p \;\av{S_{z,1}}_{0} + \frac{1}{2}\av{S_{z,2}}_{0}\Big\rbrace\, \frac{\cos(t)}{t}  \;\bigg]
\nonumber\\
& =
\bar{\av{S_{z}}} + \frac{\lx^2+\ly^2}{2}\,\delta \cdot \E^{-\I\ons t} \cdot
\nonumber\\
& \phantom{=}\;
\bigg[\Big\lbrace\av{S_{z}}_{0} - p\,\frac{\lx\ly}{\lx^2+\ly^2}\Big\rbrace\, \frac{\sin(t)}{t} +
\nonumber\\
& \phantom{=}\;
-\I\phantom{\bigg[} \Big\lbrace p \;\av{S_{z}}_{0} + \frac{\lx\ly}{\lx^2+\ly^2}\Big\rbrace\, \frac{\cos(t)}{t}  \;\bigg]
\,\text{,}
\label{eq:Sz_final}
\end{align}
which consists of a constant contribution $\bar{\av{S_{z}}}$ and a small decaying portion of order $O(\delta)$.

In contrast to the transverse spin part, the power law decaying term oscillates with two frequencies $\ons$, determined by the effective magnetic field, and $\tau_{HI}^{-1}$, which is set by the strength of the HI. The amplitude of this oscillation is decaying similarly to the corresponding part of $\av{S_{+}}(t)$ with a power law $\sim t^{-1}$. As a consequence of this, the longitudinal spin component settles to a constant value $\bar{\av{S_{z}}}$ for longer times, which is up to small corrections of order $\delta$ given by the initial value.

Note that all results are valid for arbitrary coupling constants $\lx$, $\ly$ and $\lz$, which particularly implicates, that these findings are true for a graphene QD with both a perpendicular and a parallel magnetic field. A possible negative sign of the longitudinal coupling constant $\lz<0$, as it is the case for a parallel field, can be handled by changing the sign of the magnetic fields, while using $|\lz|$, which is allowed by symmetries of $\av{S_{z,1/2}}(t)$ described in Eq. (\ref{eq:Spmz_rel}).

\subsection{Discussion of the results}
\label{subsec:comp_of_mod}

Motivated by the physics of the HI in a graphene QD, we have analyzed the effect of an anisotropic HI, where we find two distinct classes, which are characterized by $\lx=\ly$ and $\lx\neq\ly$, respectively. Representative for the former is a graphene QD subjected to a perpendicular oriented magnetic field, where $-2\lx=-2\ly=\lz=1$, but also the earlier investigated\cite{Coish2004} isotropic model for \ga{} is a member of this class. In contrast to this, a graphene QD with an in-plane magnetic field is exemplary for the second class of HI, where $\lx=-2\ly=-2\lz=1$.

From Eqs. (\ref{eq:S+_final}) and (\ref{eq:Sz_final}), one sees, that the overall dynamics of the electron spin is the same for both classes of anisotropy. The transverse components exhibit a dominant oscillating contribution, which describes a simple precession of the electron spin around the effective magnetic field consisting of the external and nuclear fields. Moreover, the transverse component features another term arising from the HI-induced flip-flop processes. Due to the large Zeeman splitting of the electron spin states, these processes are suppressed leading to a very small absolute value of this contribution compared to the precession term. Additionally, its amplitude oscillates with a much smaller frequency set by the HI and decays as a function of time, which is asymptotically well described by a power-law for times $t>\tau_{HI}$, while for even shorter times only numerical results are within reach. The longitudinal spin component exhibits a similar power law decay, whose amplitude is in addition modulated by an oscillation with the frequency corresponding to the effective magnetic field. The main part of the longitudinal spin is, however, preserved and is given up to small corrections by the initial value of the electron spin. 

While these general remarks on the electron spin dynamics are true for arbitrary anisotropy, there are, however, both qualitative and quantitative ramifications of different choices of the couplings $\lambda_{i}$, which is reflected in the amplitudes of the decaying contributions. According to Eq. (\ref{eq:Sz_final}) the longitudinal part shows only a quantitative dependence, whereas the transverse spin parts are affected quantitatively as well as qualitatively by different types of anisotropy as can be seen in Eq. (\ref{eq:S+_final}). For the first class of anisotropy characterized by $\lx=\ly$, we find that all amplitudes are proportional to the initial value $\av{S_{+}}_{0}=\av{S_{x}}_{0}+\I\av{S_{y}}_{0}$ leading to pure quantitative distinctions within this class. Since the qualitative concurrence arises from $\lx=\ly$ independently of the specific values of the constants, it seems that rather generally, an anisotropy between the transverse and longitudinal spin-subspace, does not lead to different physical behavior. In contrast to this, the second class with $\lx\neq\ly$ exhibits a $\lx$ and $\ly$ weighted mixing of the initial values $\av{S_{x}}_{0}$ and $\av{S_{y}}_{0}$. This is caused by the broken rotational symmetry the $x$-$y$ plane defining a precise reference frame according to which $\av{S_{x}}_{0}$ and $\av{S_{y}}_{0}$ can be measured.

Despite this mixing of amplitudes, one can additionally expect different timescales of the HI as is discussed in Sec. \ref{sec:key_res}. For the parallel and perpendicular case, one finds $\tau^{\parallel}_{HI}/\tau^{\perp}_{HI}=2$ due to the different coupling $\lz$ in the longitudinal direction. Thus, longer coherence times should be measurable for an in-plane magnetic field.

\section{Conclusion and Outlook}
\label{sec:conclusion}

In this article, we have shown, how the analysis of the HI by means of a generalized master equation can be extended from the isotropic case\cite{Coish2004} to a general HI, which can in principle be completely anisotropic ($\lx\neq\ly\neq\lz$). Investigating a graphene QD, with an external magnetic field applied either perpendicular or parallel to the carbon plane, we were able to study two specific realizations of an anisotropic HI, which are representative for two different classes of anisotropy. While the former system describes a situation, where the anisotropy exists between the longitudinal and transverse spin subspaces ($\lx=\ly\neq\lz$), the latter exhibits the anisotropy within the transverse subspace ($\lx\neq\ly$), where no other special relation between $\lx,\ly$, and $\lz$ is assumed. As we showed above, the first class of anisotropy gives rise only to quantitative differences leaving the analytical structure of the electron spin dynamics unchanged with respect to the isotropic case. In this model, most of the spin amplitude is preserved, while a small portion decays with a non-exponential behavior. Presumably, the reason for this universality is the fact, that the rotational symmetry within the transverse subspace is not broken for $\lx=\ly$. This symmetry is, however, not preserved for the second class of anisotropy, in which $\lx\neq\ly$, leading to a different amplitude of the power law decaying contribution. While its order of magnitude is unchanged, we find an anisotropy weighted mixing of the initial amplitude of the electron spin. As a consequence of this, its preparation with respect to the precisely defined $x$ and $y$ direction matters in contrast to a rotational symmetric system.
   
All these results were obtained for times $t\ll\Delta^{-1}\tau_{HI}$, which are determined by the ratio of the nuclear and external magnetic fields, $\Delta\propto A/\hbar \gamma_{S}B_{z}$. In this range of time, the electron nuclear spin system is in a non-Markovian regime. Considering even longer times $t\gg\Delta^{-1}\tau_{HI}$, which are not captured by our (second order) treatment, the system can, however, return to a Markovian regime again, as was found\cite{Coish2008,Coish2010} for an electron spin in a \ga{} QD. Thus, it should be interesting in the future to study how the time evolution of a spin system with anisotropic HI behaves for longer times.

Our findings are qualitatively valid for other systems fulfilling the requirements of our model, where the most important demands are a Gaussian-like envelope function, slow dynamics of the nuclear bath and a sufficiently large Zeeman-splitting with respect to the HI energy scale. Graphene-specific quantitative changes of our results arise from the small HI coupling constant A and the low natural abundance $n_{I}\approx0.01$ of $^{13}C$ leading to a prolonged time scale of the HI and a reduced nuclear magnetic field, which make the design of qubits less challenging, since computation-cycles can last for longer times and lower external fields are sufficient. These latter effects would become even more important in isotopically purified samples featuring $n_{I}<0.01$. However, if one increasingly reduces the amount of $^{13}C$ in the graphene QD, one will reach a regime, where it is not appropriate anymore to think of a bath of nuclear spins. Unfortunately, this regime is not accessible by our treatment calling for successive studies, which deal with a finite number of nuclear spins\cite{Schliemann2002,Schliemann2003}, where in particular a more detailed study of the influence of the initial conditions should be possible.

\subsection*{Acknowledgement}

We acknowledge support from the Priority Program 1459 ''Graphene`` of the DFG and from the EuroGRAPHENE Program of the ESF. Furthermore we would like to thank Bill Coish, Patrik Recher, and Stefan Walter for valuable discussions.

\appendix

\section{Calculation of the self-energy in second order}
\label{app_sec:selfen}

In this appendix we discuss the computation of the self-energy matrix-elements $\Sigma^{(2)}_{ij}$ given in Eqs. (\ref{eq:Sigma_uu}) to (\ref{eq:Sigma_mm}) in more detail. We assume, that the nuclear magnetic field operators $\gpm$ for a specific form of the anisotropy are already inserted. Thus, all parts of the self-energy are linear combinations with $\lambda_{i}$-dependent prefactors, where the summands contain one super-operator $\Fud$ or $\Gud$, two bare nuclear magnetic field operators $\hpm$ and the nuclear initial state $\ri=\ket{n}\bra{n}$. Calculating expectation values with respect to the nuclear state $\ket{n}$, all summands featuring squared operators $\hpm^{2}$ vanish identically, which reduces the number of contributions. In the following, we will present, how the remaining expressions can be evaluated in a general approach. For simplicity, we will neglect the prefactors in the following presentation.

Due to the linearity and the cyclicity of the trace, all self-energy parts in second order can be written as a linear combination of terms of the form
\begin{equation}
\Tr_{I}\left(\_\;\_\;\_\;\ri\right)
\,,
\label{app_eq:TRI_gen}
\end{equation}
where the free spots can be filled with one raising operator $\hp$, one lowering operator $\hm$ and one of the four super-operators $\Fud(\cL_{+})$ and $\Gud(\cL_{-})$ given in Eqs. (\ref{eq:Fud}) and (\ref{eq:Gud}), respectively. Altogether, this leads to $3\cdot2\cdot4=24$ possible combinations. This number can be reduced by using the relation\cite{Fick1990} of Liouvillian-like operators:
\begin{equation}
\Tr_{I}\left(f[\cL_{\pm}]\oO_{1}\oO_{2}\right)=\Tr_{I}\left(\oO_{1}f[\pm\cL_{\pm}]\oO_{2}\right)
\label{app_eq:Gua_rel}
\end{equation}
for arbitrary operators $\oO_{1,2}$ and a function $f[\cL_{\pm}]$, which can be expanded in powers of (anti-)commutators $\cL_{-}$ ($\cL_{+}$). Therefore, the result for $\Fud\left(\cL_{+}\right)$ is independent from the position, at which these super-operators are placed, while moving $\Gud\left(\cL_{-}\right)$ to a neighboring position involves a change of sign of its argument $\cL_{-}\rightarrow-\cL_{-}$. As a consequence of this, it is sufficient to calculate the following traces $\Tr_{I}(\Fud\hpm\hmp\ri)$ and $\Tr_{I}(\Gud\hpm\hmp\ri)$, which give rise to sums over expectation values with respect to  nuclear eigenstates $\ket{q}=\bigotimes_{k}\ket{m^{q}_{k}}$: 
\begin{align}
& \phantom{=} \;
\Tr_I\Big(\Fi\hpm\hmp\ri\Big)
\nonumber\\
& =
\sum\limits_{p,q,r} \bra{p}\Fi \Big\lbrace \ket{q} \bra{q}\hpm\ket{r} \bra{r}\Big\rbrace  \hmp\ket{n}\braket{n}{p}
\nonumber\\
& =
\sum\limits_{q,r} \bra{n}\Fi \Big\lbrace \ket{q} \bra{r} \Big\rbrace \bra{q}\hpm\ket{r} \hmp\ket{n}
\nonumber\\
& =
\sum\limits_{q,r} \Big[\Fi\Big]_{qr} \bra{q}\hpm\ket{r} \braket{n}{q} \bra{r} \hmp\ket{n}
\nonumber\\
& \equiv
\sum\limits_{r} \Big[\Fi\Big]_{nr} \Big[ \hpm \Big]_{nr} \Big[ \hmp \Big]_{rn}
\,\text{,}
\label{app_eq:self_element}
\end{align}
and analogously for the super-operators $\Gi$. Using Eqs. (\ref{eq:Fud}) and (\ref{eq:Gud}), one can calculate the expectation values of these super-operators:
\begin{align}
& \phantom{=}\;
\Big[\Fud\left(\beta\cL_{+}\right)\Big]_{nr}
\nonumber\\
& =
\Big\lbrace
s-\I \alpha_{\ua,\da}\beta \; \Big[b_{S}-b_{I} + \frac{\lz}{2} \sum_{k} A_{k} \left(m^{n}_{k}+m^{r}_{k}\right)\Big] 
\Big\rbrace^{-1}
\,,
\label{app_eq:Gpm_exp_val}
\\
& \phantom{=}\;
\Big[\Gud\left(\beta\cL_{-}\right)\Big]_{nr}
\nonumber\\
& =
\Big\lbrace
s+\I\alpha_{\ua,\da}\beta \; \frac{\lz}{2} \sum_{k} A_{k} \left(m^{n}_{k}-m^{r}_{k}\right)
\Big\rbrace^{-1}
\,,
\label{app_eq:Gua_exp_val}
\end{align}
where $\alpha_{\uparrow,\downarrow}=\pm1$ and $\beta=\pm1$. The factor $\lz$ is the anisotropy coefficient in $z$-direction. Next, we calculate the magnetic field expectation values, where we begin with the action of a local operator $\hat{I}_{k,\pm}$ at an arbitrary site $k$ in order to simplify later steps:
\begin{align}
\hat{I}_{k,\pm} \bigotimes_{l} \ket{m^{q}_{l}}
& =
\sqrt{I(I+1)-{m^{q}_{k}(m^{q}_{k}\pm1)}} \ket{m^{q}_{k}\pm1}
\nonumber\\
& \phantom{m}\;
\otimes \bigotimes_{l\neq k} \ket{m^{q}_{l}}
\nonumber\\
& \equiv
M_{\pm}(m^{q}_{k})\;\; \ket{m^{q}_{k}\pm1} \bigotimes_{l\neq k} \ket{m^{q}_{l}}
\,,
\label{app_eq:localrasingop}
\end{align}
We introduced the shorthand notation $M_{\pm}(m^{q}_{k})$, which obeys the relation $M_{\pm}(m^{q}_{k}\mp1)=M_{\mp}(m^{q}_{k})$. Note, that by the action of $\hat{I}_{k,\pm}$ only one single local state was changed while all other local states remain unchanged. Using the above equation, one can calculate the expectation value of the nuclear magnetic field operators $\hpm$ with respect to eigenstates $\ket{p}$, $\ket{q}$ of the $\hat{h}_{z}$-component:
\begin{align}
& \phantom{=}\;
\bra{p}\hpm \ket{q} = \Big[\hp\Big]_{pq} =
\nonumber\\
& =
\bra{p}\sum_{k} A_{k}\hat{I}_{k,\pm} \bigotimes_{p} \ket{m^{q}_{p}}
\nonumber\\
& =
\sum_{k} A_{k} M_{\pm}(m^{q}_{k})\; \bra{p} \Big\lbrace \ket{m^{q}_{k}\pm1} \otimes \bigotimes_{p\neq k} \ket{m^{q}_{p}}\Big\rbrace 
\nonumber\\
& =
\sum_{k} A_{k} M_{\pm}(m^{q}_{k})\; \Big\lbrace \bigotimes_{l'} \bra{m^{p}_{l'}}\Big\rbrace \Big\lbrace  \ket{m^{q}_{k}\pm1} \otimes \bigotimes_{l\neq k} \ket{m^{q}_{l}}\Big\rbrace 
\nonumber\\
& =
\sum_{k} A_{k} M_{\pm}(m^{q}_{k})\; \delta_{m^{p}_{k},m^{q}_{k}\pm1} \prod_{l,l'\neq k} \delta_{m^{p}_{l'},m^{q}_{l}}
\,.
\label{app_eq:nuc_mag_exp_val}
\end{align}
This equation totally sets the relation between the two sets of product states $\left\lbrace m^{p/q}_{l} \right\rbrace_{l=1}^{N}$. Inserting this result in Eq. (\ref{app_eq:self_element}), we find:
\begin{align}
& \phantom{=}\;
\sum\limits_{r} \Big[\Gud\left(\beta\cL_{-}\right)\Big]_{nr} \Big[\hpm\Big]_{nr} \Big[\hmp\Big]_{rn}
\nonumber\\
& =
\Big\lbrace
s+\I\alpha_{\uparrow\downarrow}\beta\frac{\lz}{2} \sum_{k_{1}} A_{k_{1}} \left(m^{n}_{k_{1}}-m^{r}_{k_{1}}\right)
\Big\rbrace^{-1}
\nonumber\\
& \; \times
\Big\lbrace
\sum_{k_{2}} A_{k_{2}} M_{\pm}(m^{r}_{k_{2}})\; \delta_{m^{n}_{k_{2}},m^{r}_{k_{2}}\pm1} \prod_{l,l'\neq k_{1}} \delta_{m^{n}_{l'},m^{r}_{l}}
\Big\rbrace
\nonumber\\
& \; \times 
\Big\lbrace
\sum_{k_{3}} A_{k_{3}} M_{\mp}(m^{n}_{k_{3}})\; \delta_{m^{r}_{k_{3}},m^{n}_{k_{3}}\mp1} \prod_{\tilde{l},\tilde{l}'\neq k_{3}} \delta_{m^{r}_{\tilde{l}'},m^{n}_{\tilde{l}}}
\Big\rbrace
\nonumber\\
& =
\sum_{k_{2}}\frac{A^{2}_{k_{2}} M^{2}_{\mp}(m^{n}_{k_{2}})}{s\pm\I\alpha_{\uparrow\downarrow}\beta\frac{\lz}{2}A_{k_{2}}}
\,.
\label{app_eq:self_element_Gua}
\end{align}
The remaining two equations are obtained in the same manner leading to:
\begin{align}
& \phantom{=}\;
\sum\limits_{r} \Big[\Fud\left(\beta\cL_{+}\right)\Big]_{rn} \Big[\hpm\Big]_{nr} \Big[\hmp\Big]_{rn}
\nonumber\\
& =
\sum_{k_{2}}\frac{A^{2}_{k_{2}} M^{2}_{\mp}(m^{n}_{k_{2}})}{s-\I\alpha_{\ua,\da}\beta\left(\ons \pm \frac{\lz}{2}A_{k_{2}}\right)}
\,,
\label{app_eq:self_element_Gpm}
\end{align}
where we used, that
\begin{align}
& \frac{\lz}{2}\sum_{k'} A_{k'} \left(m^{r}_{k'}+m^{n}_{k'}\right)\prod_{l\neq k}\delta_{m^{r}_{l},m^{n}_{l}}\delta_{m^{r}_{k},m^{n}_{k}\mp1}
\nonumber\\
& =\lz\av{\hz}_{n}\mp\frac{\lz}{2}A_{k}
\,.
\end{align}
The functions $M^{2}_{\pm}(m^{n}_{k})$ over nuclear eigenvalues can be replaced by their average assuming a nuclear state which is highly degenerate\cite{Coish2004,Barnes2011}:
\begin{equation}
\av{\av{M^{2}_{\pm}(m^{n}_{k})}}=\cpm
\,\text{.}
\label{eq:cpm}
\end{equation}
Finally, for a large nuclear spin system with $N\gg1$, the remaining sums in Eqs. (\ref{app_eq:self_element_Gua}) and (\ref{app_eq:self_element_Gpm}) can be replaced by integrals in the continuum limit. Changing to dimensionless units by measuring energies in units of $\epsilon_{HI}=|\lz|n_{I} A/2N$, one finds up to small corrections\cite{Coish2004}:
\begin{align}
& \phantom{=}\;
\Tr_I\Big(\Fud\left(\beta\cL_{+}\right)\hpm\hmp\ri\Big)
\nonumber\\
& =
\frac{4 N}{\lz^{2}} \cmp I_{\mp\left\lbrace\alpha_{\ua\da}\beta\tlz\right\rbrace}(s-\I\alpha_{\ua\da}\beta\ons) 
\label{app_eq:Gua_final}
\\
\nonumber\\
& \phantom{=}\;
\Tr_I\Big(\Gud\left(\beta\cL_{-}\right)\hpm\hmp\ri\Big)
\nonumber\\
& =
\frac{4 N}{\lz^{2}} \cmp I_{\mp\left\lbrace \alpha_{\ua\da}\beta\tlz\right\rbrace}(s)
\label{app_eq:Gpm_final}
\,,
\end{align}
where $\tlz=\lz/|\lz|$ and
\begin{equation}
I_{\pm}\left(s\right)
=
s\left[\log\left(s\mp \I\right)-\log\left(s\right)\right]\pm \I
\,\text{.}
\end{equation}
Applying this continuum limit, however, sets an upper bound $t\ll\sqrt{N/2}\tau_{HI}$ as discussed in the main text. Knowing the four basic terms given in Eqs. (\ref{app_eq:Gua_final}) and (\ref{app_eq:Gpm_final}), respectively, all other remaining possible summands to the self-energy are readily obtained by using Eq. (\ref{app_eq:Gua_rel}).

\end{document}